\documentclass[a4paper,oneside]{article}

\usepackage[dvips]{graphicx}

\usepackage{amsmath}
\usepackage{amsfonts}
\usepackage{amssymb}
\usepackage{amsthm}
\usepackage{amscd}
\usepackage[mathscr]{eucal}

\usepackage{hyperref}

\newcommand{\myURL}[1]{\url{#1}}

\newcommand{\myRef}[1]{
  \S\ref{#1}
}

\newcommand{\etc}{\textit{etc.}}
\newcommand{\eg}{\textit{e.g.}\,}
\newcommand{\ie}{\textit{i.e.}\,}

\newcommand{\etal}{\textit{et al.}\,}

\newcommand{\viz}{\textit{viz.}\,}

\newcommand{\Java}{\emph{Java}\,}

\newlength{\facewd} \newlength{\faceht}

\theoremstyle{plain}
\newtheorem{theorem}{Theorem}[section]

\theoremstyle{definition}

\newtheorem{scenario}{Scenario}[section]
\newtheorem{condition}{Condition}[section]
\newtheorem{definition}{Definition}[section]
\newtheorem{notation}{Notation}[section]
\newtheorem{remark}{Remark}[section]

\newcommand{\droptext}[1]{\ensuremath{
    \text{\ #1\ }}
  }

\newcommand{\field}[1]{\ensuremath{\Bbb{#1}}}

\newcommand{\eqdef}{\ensuremath{\triangleq}}

\newcommand{\powerSet}[1]{\ensuremath{
    \mathscr{P}\left(#1\right)
    }
  }

\newcommand{\myR}{ \ensuremath{ \mathrel{R} } }

\newcommand{\myVec}[2]{\ensuremath{
    \Vec{#1}^{\langle #2 \rangle }
    }
  }

\newcommand{\dset}{\ensuremath{
    \{ 1, 0, -1 \}
    }
  }

\newcommand{\sign}{\ensuremath{
    \operatorname{\textbf{s}}
    }
  }

\newcommand{\promote}{\ensuremath{
    \operatorname{promote}
    }
  }

\newcommand{\demote}{\ensuremath{
    \operatorname{demote}
    }
  }

\newcommand{\sincere}{\ensuremath{
    \operatorname{sincere}
    }
  }

\newcommand{\sophis}{\ensuremath{
    \operatorname{sophisticated}
    }
  }

\newcommand{\versusOp}{\ensuremath{
    \mathrel{\operatorname{vs}}
    }
  }

\begin{document}

\title{Collective Choice Theory in Collaborative Computing}

\author{Walter D Eaves}


\date{\today}

\maketitle

\begin{abstract}
  This paper presents some fundamental collective choice theory for
  information system designers, particularly those working in the field of
  computer--supported cooperative work. This paper is focused on a
  presentation of Arrow's Possibility and Impossibility theorems which form
  the fundamental boundary on the efficacy of collective choice: voting and
  selection procedures. It restates the conditions that Arrow placed on
  collective choice functions in more rigorous second--order logic, which
  could be used as a set of test conditions for implementations, and a
  useful probabilistic result for analyzing votes on issue pairs. It also
  describes some simple collective choice functions. There is also some
  discussion of how enterprises should approach putting their resources
  under collective control: giving an outline of a superstructure of
  performative agents to carry out this function and what distributing
  processing technology would be needed.
\end{abstract}

\section{Collective Choice in Information Systems}
\label{sec:choice}

\subsection{Naming Services}

\begin{enumerate}
  
\item \textit{Windows NT}
  
  If one uses a system then, from time to time, one might receive an event
  message saying that ``The Browser Has Forced an Election...'' %
  \cite{election:microsoft:browser:msg,election:microsoft:browser}.
  
\item The Internet and the Domain Naming Service DNS
  
  Internet connected systems could not function without the Domain Naming
  Service and this too relies upon elections: individual system
  administrators choose when their name--server is to authoritative or not %
  \cite{sec:dnssec-intro,dns:setup} and which name--servers it will rely upon.

\end{enumerate}

\begin{figure}[htbp]
  \begin{center}
  \includegraphics{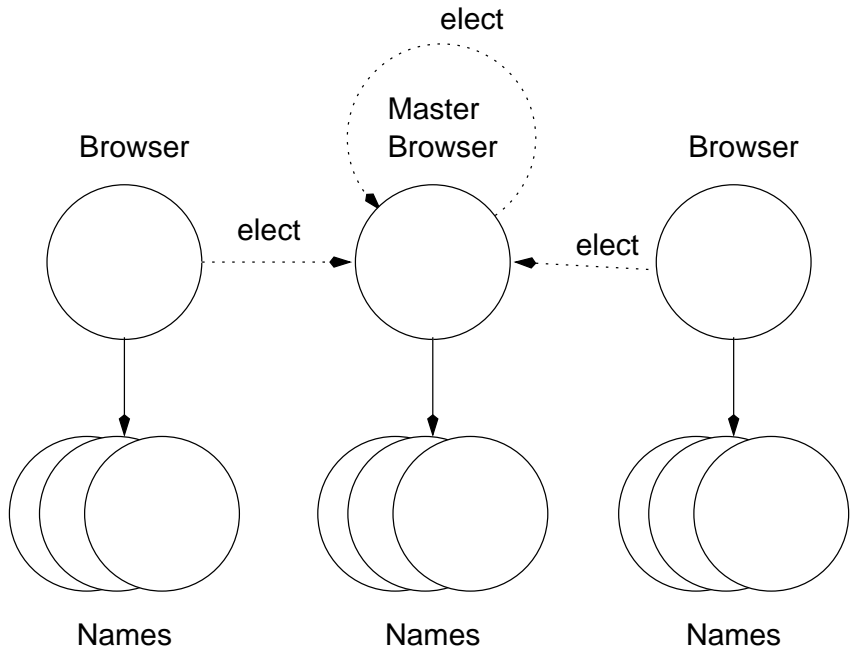}
  \caption{Browser programs choosing a master browser}
  \label{fig:elect-browser}
  \end{center}
\end{figure}

The difference between the two systems is that \textit{Windows NT} is
designed to manage the naming of relatively small domains and can use a
direct election amongst all its naming components, the browsers. This is
represented in figure \ref{fig:elect-browser}, where the master browser is
elected by the itself and the other browsers.

\begin{quotation}
  The browsers apply the same fitness criteria in choosing their
  master. There can be no conflict in policy.
\end{quotation}

\begin{figure}[htbp]
  \begin{center}
  \includegraphics{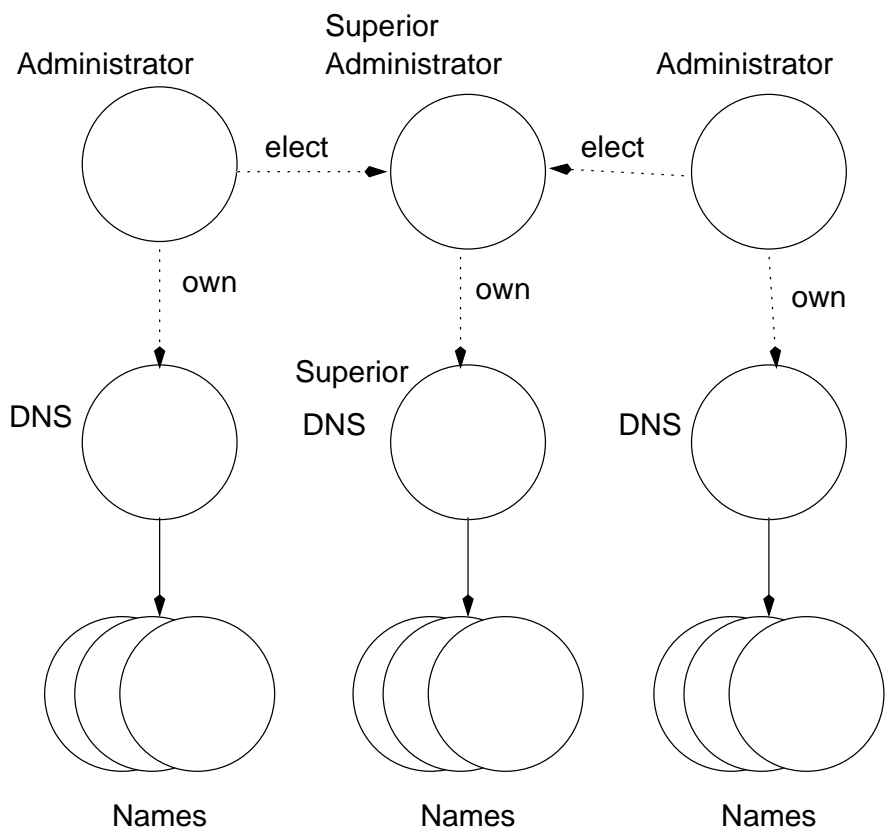}
  \caption{DNS Administrators choosing a superior DNS}
  \label{fig:elect-dns}
  \end{center}
\end{figure}

The Internet's DNS has to rely upon a loosely co--ordinated database of
name servers. This is represented in figure \ref{fig:elect-dns}. Here the
Administrator of each DNS chooses which other DNS it will use to resolve
names. In this example, for all names other than their own, the Superior
Administrator is chosen as authoritative.

\begin{quotation}
  The Administrators need not apply the same fitness criteria in choosing
  their superior DNS. There may be conflicts in policy.
\end{quotation}

They can therefore attempt malevolent actions collectively if they so wish,
for example:

\begin{quotation}
  Consider an electronic commerce web site. The user's web browser makes a
  secure connection to the site, providing a protected channel. If the DNS
  entry for the server's address was replaced by one indicating an
  attacker's address, the browser will connect to the malicious site,
  possibly without the user's knowledge. In this scenario, the DNS spoofer
  could monitor the traffic over the ``secure'' connection, since the
  secure connection would actually be to the spoofer, and forward the
  transaction data to the real website or process the traffic itself
  \cite{sec:dnssec-intro}.
\end{quotation}

Without human intervention, computer programs have wholly predictable
behaviour and cannot possess ulterior motives: people can. (The problems
that can arise from badly managed networks are described to a much greater
extent in \cite{sec:survivable}.)

\begin{quotation}
  \emph{But} even if computer programs have predictable behaviour they may
  not apply the same fitness criteria in making choices. This may lead to
  conflicts in policy.
\end{quotation}

\subsection{Groups choosing policies}

Most people will have come across moderated newsgroups and mailing lists.
Potentially these services could be made self--managing and would be simple
examples of a computer--supported working environments.

\begin{enumerate}
\item Joining the Group
  
  A policy decision is needed to determine whether or not an individual
  should be allowed to join a particular group and take the rights and
  privileges enjoyed by its members. A collective choice, probably by a
  membership committee, is made based upon the applicant's credentials.
  
  The procedure is usually carried out using a set of recommendations. The
  individual requesting the rights fills in a form stating his credentials.
  People are assigned to check the applicant's trustworthiness,
  qualifications and so forth. If the membership committee is satisfied,
  someone is instructed to assign the applicant to the group.

\item Expulsion from the Group

  Should the membership committee decide to expel an individual from the
  group, that, too, would be a collective policy decision.

\end{enumerate}

A situation that could clearly arise is for a number of individuals to
infiltrate a group, subvert it by having themselves elected to the
membership committe and then expelling all the members of the group who are
not sympathetic to the infiltrators. Of course, those individuals might
also do this legitimately, if the selection of the membership committee
reflected the views of the current membership. What legitimates actions is
a wider consensus.

\subsection{Lattices and Access Control}

Organizing the membership of groups is a sub--process needed for the
formation of lattices of membership classes for an access control system,
first put forward by Denning in \cite{Denning:1976:LMS}. The only provably
safe access control systems are those that are based on Mandatory Access
Control, MAC, schemes, also as described by Denning \cite{sec:denning}.

\subsubsection{Operating Systems}

Discretionary Access Control, DAC, schemes are used in most multi--user
operating systems such as \textit{VAX--VMS} \cite{VAX84} and \textit{Unix}
\cite{Curry9004:Improving} and resource--sharing operating systems such as
\textit{Windows NT} \cite{sec:nt}. These are not as discretionary as one
might think:
\begin{itemize}
\item Individual users are allocated to groups
\item Privileged User(s): the ``super--user'' or ``Administrator'' set
  group memberships.
\end{itemize}

It could be described as a dictatorial Discretionary Access Control scheme.
The only latitude that individuals have is to be able to grant or deny
access rights to members of their own group or to everyone. The privileged
user can undo any access control operations performed by any individual. 

To use some better terminology: \emph{subjects} are entities who may
possess access rights and \emph{objects} are those entities to which
subjects have rights to use. An operating system that uses a MAC scheme is
represented in figure \ref{fig:elect-mac} and one that uses a DAC scheme is 
represented in figure \ref{fig:elect-dac}.

\begin{enumerate}
\item MAC Scheme

  This is a simple scheme. The administrator classifies all the subjects
  and the objects and it classifies some subjects above other subjects so
  that higher subjects can access everything that lower subjects can.

  \begin{figure}[htbp]
    \begin{center}
      \includegraphics{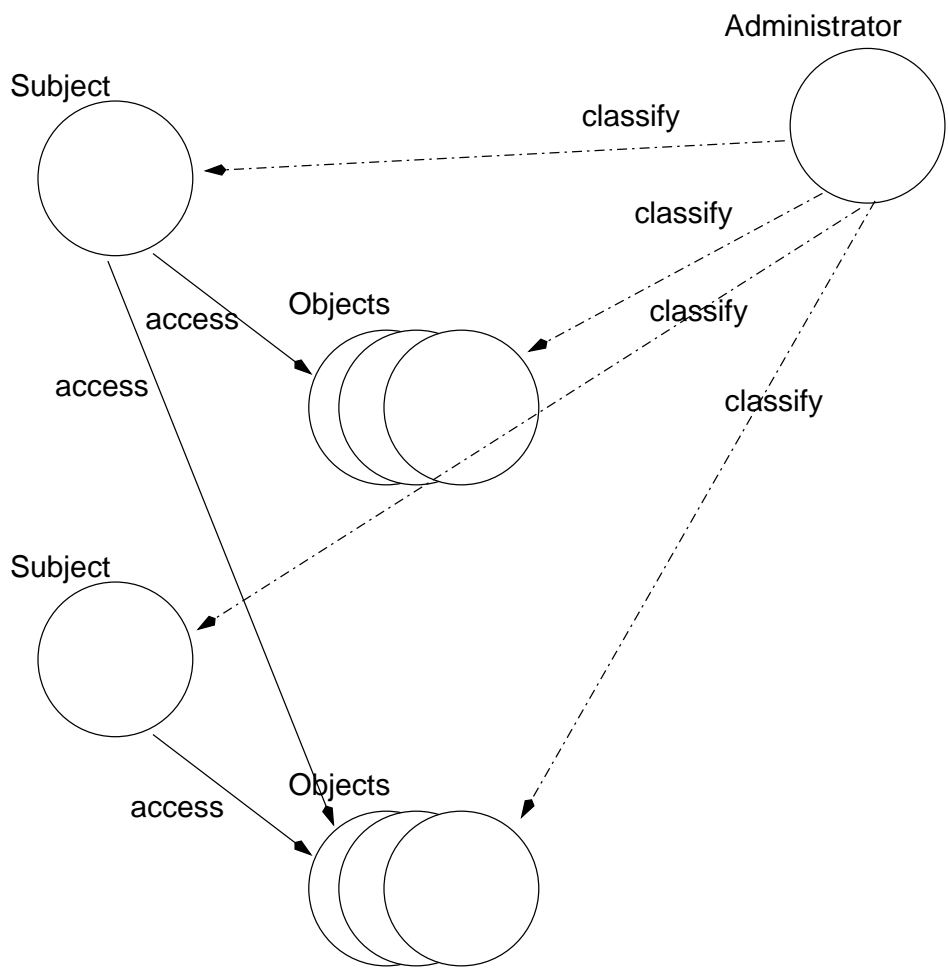}
      \caption{Mandatory Access Control Scheme}
      \label{fig:elect-mac}
    \end{center}
  \end{figure}

\item DAC Scheme
  
  This is more sophisticated. The administrator classifies all the subjects
  and can denote that they belong to certain a \emph{Group}. Every subject
  belongs to the group of \emph{Everyone}.

  Individual subjects own some objects and can choose to grant access to
  \begin{itemize}
  \item Either: their groups (or groups)
  \item Or: to \emph{Everyone}
  \end{itemize}
  
  \begin{figure}[htbp]
    \begin{center}
      \includegraphics[angle=-90,totalheight=6in]{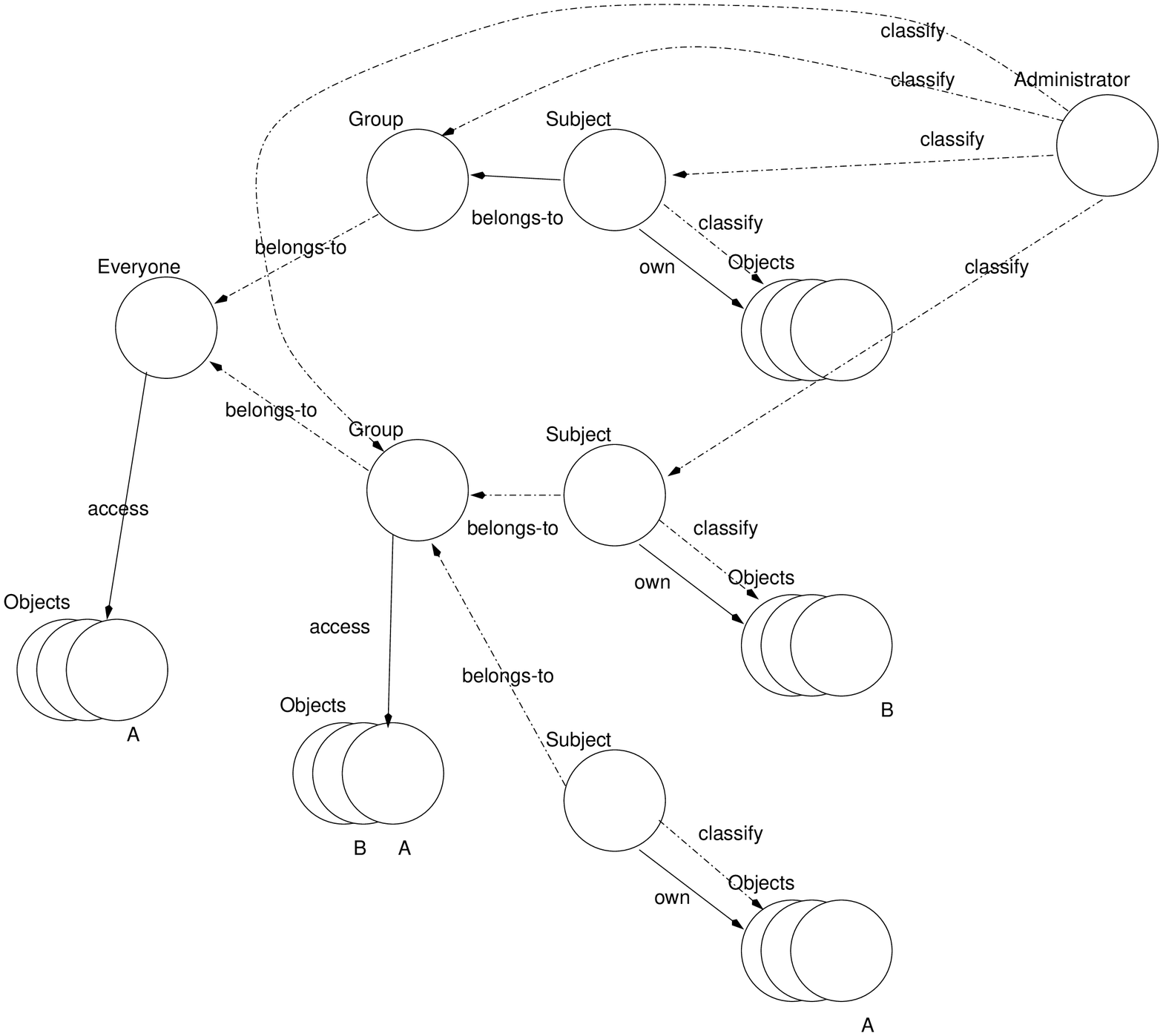}
      \caption{Discretionary Access Control Scheme}
      \label{fig:elect-dac}
    \end{center}
  \end{figure}

  Groups do not own anything and neither can the group \emph{Everyone}. In
  figure \ref{fig:elect-dac}, the object $A$ is owned by one of the
  subjects and has allowed access to the group its owner belongs to and to
  \emph{Everyone}. The object $B$ can be accessed by the group, but not by
  \emph{Everyone}.
  
  There is degree of autonomny granted to the subjects in that they may
  classify objects to be accessible to the members of the groups they
  belong to, but they may not choose which group, or groups, they belong
  to. Neither may they change ownership of an object they own\footnote{A
    \emph{POSIX}\cite{Lewine:1991:PPG} requirement, \textit{Unix} typically
    does allow ownerships to be changed.}.
  
\end{enumerate}

\subsubsection{Database Management Systems DBMS}

Some DBMSs have more flexible DACs which allow some individuals to be more
privileged than others by granting them the right to grant rights. This
feature is available in some DBMSs that support the Structured Query
Language, SQL, which was based upon \textit{System--R}, which is descibed
by Denning in \cite{sec:denning:2}.

In effect, this is the same as the DAC scheme for operating systems, see
figure \ref{fig:elect-dac}, but the owner can grant access to subjects
other than those in its group and it can grant to others the right to grant
access, but not to those to whom access has been explicitly denied by the
owner.

\subsection{General Resource Management Systems}
\label{sec:gen-access}

Information processing systems can be thought of as general resource
management systems and the Open Distributed Processing ODP standards, in
particular the prescriptive model, \cite{ODP:presc}, describe how an
enterprise modelling language could be used to state the relationships
between resource owners and users as behavioural contracts given in terms
of a set of permissions, prohibitions and obligations. This can be seen as
a generalization of the MAC and DAC schemes, but the ODP standards are only
reference models and each information system should contain some component
that embodies its enterprise model. This is a wholly new superstructure to
an information processing system: there are some simple class relationships
describing the ODP model set out in appendix \ref{sec:odp-enterprise},
these rather vaguely state the information model for the
superstructure. Some mechanisms that could be used for applying policy,
implementing the administrators, has already been proposed
\cite{ch-o:eaves}.

Recently, the \Java programming environment has provided a powerful
language for expressing permissions \cite{sec:java:security}. However, it
assumes the existence of some agent which would negotiate the contracts
between resource owner and user. There have been only a few efforts made to
develop arbitrator agents which could generate such contracts.

There are some system proposals which attempt to apply abstract behavioural
rules in terms of concrete permissions: in papers by Minsky
\cite{minsky89a,db:pol:minsky}. Minsky's treatment is for information
processing systems in general, but a paper by Rabitti \etal
\cite{Rabitti:1991:MAN} describes authorization generation mechanisms which
support a lattice model of authorization policy for an object--oriented
database. The innovation of the system is that it generates authorization
policy as it operates.  Authorization is viewed as having three dimensions:
 
\begin{description}
\item[Expression] Authorizations specified by users, which are known as
  \emph{explicit} and those that are derived by the system as known as
  \emph{implicit}.
\item[Direction] An authorization can be \emph{positive}, stating what may
  be done, or \emph{negative} stating what may not be done.
\item[Strength] An authorization may be \emph{strong}, in which case it may
  not be overridden, or \emph{weak}, in which it can.
\end{description}

This model has been extended \cite{db:pol:bertino} and a recent
contribution by Castano \cite{db:pol:castano} introduces metrics that can
be used to generate concrete permissions from more abstract specifications,
including:
\begin{itemize}
\item Operation compatibility
\item Individuality similarity co--efficient
\item Authorization compatability
\item Semantic correspondence
\item Clustering of Individuals
\end{itemize}

Although Bertino \etal attempt to produce mechanical means of generating
authorization policy no--one would seriously expect a system to be driven
wholly by mechanical recommendation, it would require choices to be made by
people and, if that is the case, then a suitable collective choice
procedure must be found.

\subsection{Summary}

Information processing systems are not mechanical systems. They represent
the interests of people and the agents that comprise an information
processing system will require policies that represent people's interests
which will be used in formulating behavioural contracts between agents that
own resources and the agents that use them.

\section{Issues in Collective Choice Theory}

The principal difficulty in collective choice theory is that if a group of
people have to choose between more than two issues, there is no choice
procedure they can adopt that is \emph{not} open to abuse. By abuse, it is
meant:
\begin{itemize}
\item Denial of service: some agent can exercise a veto on any policy
  proposed.
\item Enforced service: some agent can force a policy upon others.
\end{itemize}
An abuse takes the form of insincere behaviour: an agent acts not to fulfil 
his own interests, but to prevent others from fulfilling theirs.

This simple example, known as the \textit{Voting Paradox} might help:

\begin{scenario}[Denial of Service by Policy Cycle] Three agents, $x$, $y$
  and $z$, have to choose between three services: $A$, $B$ and $C$. $x$ and
  $y$ rank sincerely, but $z$ ranks the policies to prevent $x$ and $y$
  from reaching a compromise, \ie policy $A$, see table \ref{tab:loop}.
  \begin{table}[htbp]
    \begin{center}
      \begin{tabular}[left]{|l|r|}
        \hline
        Agent & Ranking \\
        \hline
        $x$ & $A>B>C$ \\
        $y$ & $C>A>B$ \\
        $z$ & $B>C>A$ \\
        \hline
      \end{tabular}
      \caption{Policy cycle used to deny service}
      \label{tab:loop}
    \end{center}
  \end{table}
\end{scenario}

Just to clarify terminology: an election is an expression of collective
choice and the policy chosen by an election is the outcome of what
statisticians might call a trial. In a trial, there are a number of choices
available to each voter, or individual, taking part. A policy is usually
chosen with regard to an issue; the proposal that a policy should be
followed regarding an issue is called a motion.

\subsection{Two Policy Issues}

This could also be described as a a two outcome trials. These arise when
there are only two policies which can result: the choice is to accept a
policy or not.

\begin{enumerate}
\item Number of Choices is always three
  
  Although there are two policies, there are three choices: one can vote
  \emph{For} or \emph{Against}. One may also be given the explicit right to
  \emph{Abstain}, and, by doing so, state that one cannot vote for or
  against. One may also choose \emph{Not to Vote}.
  
\item Abstentions and Not Voting
  
  Usually in referenda, there is no option to abstain and those who do not
  vote are considered to have abstained. This assumption is legitimate if
  one is sure that all individuals who are entitled to vote have been
  informed that they may do so and have made a decision not to.  In most
  business processes, this would not be the case.
  
  In what follows, it is assumed that all abstentions are explicitly made
  and that individuals who do not vote have excluded themselves.
  
\item Choice Function
  
  Simple majority rule is the usual method for resolving two policy
  elections, but it by no means the only one. Two policy collective choice
  is dealt with in some detail in appendix \ref{sec:scf}.

\end{enumerate}
  
What is required from a choice function is that it is not open to abuse and 
it is decisive. Two policy issues can not be abused because it can be shown 
to be the case that the majority have chosen the policy. The only
difficulty is resolving ties.

\subsection{Three policy issues}
\label{sec:three-policy}

A three--outcome trial, for example, one of $A, B$ or $C$ \emph{must} be
chosen. The voting procedures described here are explained at greater
length in \cite{vote:saari} and and there is some documentation on voting
methods at \cite{vote:qub}. What follows illustrates some of the problems
that arise from using them.

\begin{enumerate}
\item Number of Choices can vary

  \begin{enumerate}
  \item Four Choices
    
    One can organize the election so that are four choices: any of $A, B$
    or $C$ and to abstain.
    
  \item Seven (or Eleven) Choices
    
    One could also allow voters to express a choice between their first two
    preferences and to abstain. So a vote might be: $(A, B)$ which implies
    that $A>B>C$.
    
    One can also allow voters to state they are indifferent between their
    first two choices, but prefer them to the third; an example of this
    kind of vote is: $(A=B)>C$.
    
  \item Thirteen (or Seven) Choices
    
    Also one can allow voters to express their choices as a ranking over
    all three policies, in two ways:
    \begin{itemize}
    \item Strong ordering \- no statements of indifference allowed.
    \item Weak ordering \- statements of indifference are allowed.
    \end{itemize}
    
    The former allows seven choices of rankings, the latter thirteen.

  \end{enumerate}
  
\item Choice Functions

  \begin{enumerate}
    
  \item Simple Majority Rule
    
    This could only be used when the voters are presented with four choices
    and clearly would not work:
    \begin{itemize}
    \item Three voters: a tie can result from a policy cycle and can be
      used by one voter to deny service, see table \ref{tab:loop}.
    \item Seven voters: if three vote $A$, two $B$ and two $C$, then, even
      though a majority did not want $A$, $A$ is chosen.
    \end{itemize}
    
  \item Single Transferable Vote
    
    This could be used when the voters are presented with seven or eleven
    choices. It suffers from the same problem as the next procedure.
    
  \item Hare Voting System and Borda Preferendum
    
    These two can be used if one presents to the voters thirteen or seven
    choices; they, and the Single Transferable Vote procedure, all suffer
    from the same fault, \cite{scf:doron,scf:fishburn:1}, which is that
    voting is affected by irrelevant alternatives. This is best illustrated
    by using the results of a Borda Preferendum which has each voter rank
    their alternatives in order, see table \ref{tab:choice:1}.

    \begin{table}[htbp]
      \begin{center}
        \begin{tabular}[left]{|l|r|r|r|r|}
          \hline
          & \multicolumn{4}{c|}{Policy and Ranking} \\
          Voter & w & x & y & z \\
          \hline
          i & 4(3) & 3(-) & 2(2) & 1(1) \\
          j & 4(3) & 3(-) & 2(2) & 1(1) \\
          k & 1(1) & 2(-) & 4(3) & 3(2) \\
          \hline
            & 9(7) & 8(-) & 8(7) & 5(4) \\
          \hline
        \end{tabular}
        \caption{The Borda ``Preferendum''}
        \label{tab:choice:1}
      \end{center}
    \end{table}
    
    If the voters $i, j$ and $k$ are asked to rank the four policies $w, x,
    y$ and $z$, then the order is $w > x > y > z$, but if asked to choose
    between $w, y$ and $z$ then $(w=y) > z$, but it was clear that $w$ was
    preferred over $y$. The anomaly being that an irrelevant policy, $x$,
    serves to differentiate between relevant ones.
    
  \item Condorcet Procedure
    
    This procedure is often employed in committees but also suffers from
    irrelevant alternatives affecting the selection of a final choice.
    (There are some good examples of how a Condorcet procedure can be
    abused in \cite{vote:saari}.) It is simply a series of two policy
    elections: $(A, B)$, $(A, C)$ and $(B, C)$. If any policy beats the
    other two, then it is chosen. If there is a policy cycle then there is
    no Condorcet winner. It may also allow an irrelevant alternative to
    beat a potential Condorcet winner.

  \end{enumerate}

\end{enumerate}

It will be shown that there is no satisfactory choice function for more
than two policy issues. The next note shows that the number of different
orderings for a given number of issues increases dramatically.

\subsection{$n$ policy issues}

The total number of different weak orderings for $n$ policies can be
calculated as follows:

\begin{enumerate}
\item Generate all the partitions\cite[p. 56]{graph:skiena} of $n$.
\item Calculate the number of permutations for each partition, call this
  $N(\operatorname{partitions})$.
\item For each partition find the number of number of ways in which the
  policies could be allocated to the elements of the partition,
  $N(\operatorname{policies})$.
\item Multiply $N(\operatorname{partitions})$ by
  $N(\operatorname{policies})$ for each partition and sum them together.
 \begin{equation}
   \label{eq:preferences}
   \Sigma_{\operatorname{partitions}}
   N(\operatorname{policies}) \
   N(\operatorname{partitions}) 
 \end{equation}
\end{enumerate}

A \emph{Mathematica}\cite{graph:wolfram} package is
available\cite{sft:aidan:math} that performs the calculation. Table
\ref{tab:orders} lists the total number of different preference orders
for up to 6 policies and clearly shows how large the search space
becomes.

\begin{table}[htbp]
 \begin{center}
   \begin{tabular}[left]{|r|r|}
     \hline
     Policies & Orderings \\
     \hline
     1 & 1 \\
     2 & 3 \\
     3 & 13 \\
     4 & 75 \\
     5 & 541 \\
     6 & 4683 \\
     \hline
   \end{tabular}
   \caption{Number of Different Preference Orderings for $n$ policies}
   \label{tab:orders}
 \end{center}
\end{table}

\subsection{Sincere and Sophisticated Voting}

Sophisticated voting utilizes some strategy whereby a voter does not vote
for their first choice to ensure that their least--preferred policy is not
chosen. For example, an electorate of seven votes sincerely for three
policies $x, y$ and $z$ thus: 4 $x$, 3 $y$ and 2 $z$, then $x$ would be
chosen. However, the $z$ voters may prefer $y$ to $x$ so their
sophisticated vote is for $y$.

An interesting example of a sophisticated vote is global abstention. If a
motion is formulated which requires a choice between $A$ or $B$, but all
voters prefer $C$ which is not proposed, a sophisticated response is a
global abstention.

Unfortunately, sophisticated voters enjoy an advantage over sincere voters,
but, to do so, they must formulate their own voting policy, which usually
requires that they have some information as to the relative strengths of
the different coalitions within an electorate and their choice of voting
policy would, presumably, be decided by a sincere vote amongst them. One of
the attractions of presenting an electorate with a complex agenda---of more
than two issues---is that they are less able to formulate strategies
amongst themselves, so that complex agendas should elicit more sincere
voting, but the likelihood of a policy cycle arising is greater, as is made
clear in a later section, \myRef{sec:max-min} and by equation
\eqref{eq:scf:max-min}. Sincere voting would allow voters' underlying
values to more precisely determined, which one would hope, would in the
long--run be a more stable basis for decision--making.

\section{Collective Choice Mathematical Model}

This is more rigorous presentation of collective choice theory. This
following section introduces the notation that will be used to formalize
the conditions that are placed on collective choice functions.

\subsection{Some Notation}
\label{sec:scf:notation}

\begin{notation}[Relations, Preferences and Their Ordering: $\succ$] is a
  preference ordering over a set of objects in a finite set $X = \{ x_1,
  \dots, x_n\}$ constructed thus:
 \begin{enumerate}
 \item[$\myR$] is an instance of a class of binary relations between any
   two objects. To state that $x_1$ is related to $x_2$ in some way, one
   would write: $x_1 \myR x_2$. The particular relation might be any of the
   following $>, \ge, \le, =$. At this stage, $\myR$ is taken to be
   transitive and connected.
 \item[$\succ$] is a statement of an \emph{individual's preference order}
   or \emph{preferences} over the elements of $X$. It is a tuple, \ie a
   vector, with exactly $\# X$ elements, with each element being of the
   form $x_i \myR x_j$, where $\myR$ is instantiated to one of the values
   that the class might take. The ordering is assumed to be consistent for
   whatever qualities $\myR$ possesses. There must be at least one
   statement of preference for each element of $X$, even if that statement
   is one of indifference. It is assumed that such a preference ordering is
   consistent with the qualities of $\myR$. The power set of $\succ$ is $X
   \times X$
 \end{enumerate}
\end{notation}

\begin{notation}[Policies and Voters] Some simple set definitions are
  needed.

  \begin{description}
  \item[$I$] is the set of \emph{voters or individuals} $I = \{ 1, \dots, i,
    \dots, n \}$.
  \item[$X$] a non--empty set is the universal set of social alternatives,
    or \emph{policies}, at least one of which must be chosen by the voters.
  \item[$\mathcal{X}$] is a subset of the power set $\powerSet{X}$ of $X$;
    it is a non--empty set of non--empty subsets of $X$ and describes the
    \emph{potential feasible policy sets} of $X$.
  \item[$Y$] is an element of $\mathcal{X}$. It is the set of policies that
    are presented to an electorate for them to vote on: the \emph{proposal
      set}.
  \item[$\Vec{D}$] is a preference profile of all voters, it will be called
    a \emph{vote}, but will contain more than just the voters' preferences
    on the elements of the proposal set. It contains the preference orders
    of all the individuals in the society for all alternatives in $X$.
    For the $n$ individuals, if individual $i$ is presumed to have a
    preference order ${\succ}_i$, $\Vec{D}$ can be written as the $n$-tuple
    $( \succ_1, \succ_2, \dots, \succ_n )$ of preference orders on $X$.
  \item[$\powerSet{\Vec{D}}$] is the power set of all votes, feasible and
    infeasible. For a given set of policies and a given set of individuals
    only a subset of these votes will occur.
  \item[$(Y,\Vec{D})$] is an ordered pair called the \emph{situation}. It
    is the feasible set of policies, $Y$, presented to the electorate, and
    a vote $\Vec{D}$.
  \end{description}
  The important word is feasible. Only some votes will be feasible given
  the preferences held by voters; therefore only some policy sets will be
  feasible.
\end{notation}

\begin{definition}[Sincere and Sophisticated] The following function
  definitions clarify how voters make up their minds and form their
  preference orders. They are, therefore, purely notional and one or the
  other is performed by each individual, $i$. How a preference order is
  formed is dependent on whether the individual votes sincerely or is
  sophisticated.
  
  If an individual, or population, is voting sincerely, then:
  \begin{equation*}
    \begin{aligned}[t]
      \sincere & \colon I \times X \mapsto X \times X \\
      \sincere & \colon X \mapsto \powerSet{\Vec{D}}
    \end{aligned}
    \droptext{\eg}
    \begin{aligned}[t]
      \succ_i & = \sincere(i, X) \\
      \Vec{D} & = \sincere(X)
    \end{aligned}
  \end{equation*}
 
  If an individual is a sophisticated voter, then a new set of histories is
  needed: $\mathcal{\Vec{D}}$---and its power set
  $\powerSet{\mathcal{\Vec{D}}}$---which is the set of all votes that have
  taken place.
  \begin{equation*}
    \begin{aligned}[t]
      \sophis & \colon I \times X \times \powerSet{\mathcal{\Vec{D}}}
      \mapsto X \times X \\
      \sophis & \colon X \mapsto \powerSet{\Vec{D}}
    \end{aligned}
    \droptext{\eg}
    \begin{aligned}[t]
      \succ_i & = \sophis(i, X, \mathcal{\Vec{D}}) \\
      \Vec{D} & = \sophis(X, \mathcal{\Vec{D}})
    \end{aligned}
  \end{equation*}
\end{definition}

\begin{definition}[Promotion and Demotion] When stating conditions it is
  useful to construct votes from other votes. These may be elements of
  $\powerSet{\Vec{D}}$ that are infeasible.

  These functions promote and demote a policy within a vote.
  \begin{equation*}
    \begin{aligned}[t]
      \promote & \colon \powerSet{\Vec{D}}
      \times X \mapsto \powerSet{\Vec{D}}
    \end{aligned}
    \droptext{\eg}
    \begin{aligned}
      \Vec{D}' & = \promote(x, \Vec{D})
    \end{aligned}
  \end{equation*}
  And similarly,
  \begin{equation*}
    \begin{aligned}[t]
      \demote & \colon \powerSet{\Vec{D}} \times X \mapsto \powerSet{\Vec{D}}
    \end{aligned}
    \droptext{\eg}
    \begin{aligned}
      \Vec{D}' & = \demote(x, \Vec{D})
    \end{aligned}
  \end{equation*}

  $\demote()$ would be implemented as follows:
  \begin{enumerate}
  \item Every preference not involving $x$ is unchanged:
    \begin{equation*}
      \begin{split}
        \forall x', y' \in X & [ (x' \ne x,\ y' \ne x, \ x' \myR y' \in
        \succ_i,\ 
        x' \myR' y' \in {\succ'}_i) \\
        \quad & \rightarrow (x' \myR y' \leftrightarrow x' \myR' y') ]
      \end{split}
    \end{equation*}
  \item Everything that involves $x$ is unchanged if $x$ if preferred over
    something else; or is changed so that $x$ is now preferred over the
    other policy
    \begin{equation*}
      \begin{split}
        \forall y' \in X &
        [  x > y' \in \succ_i \rightarrow x > y' \in {\succ'}_i ] \\
        \droptext{or} & \\
        \forall y' \in X & [ x = y' \in \succ_i \rightarrow x > y' \in
        {\succ'}_i ]
      \end{split}
    \end{equation*}
  \end{enumerate}

  Along the same lines, two other variants of $\promote$ and $\demote$
  can be defined, which promote or demote for a particular voter on a
  particular policy and resolve any conflicts.
  \begin{equation*}
    \begin{aligned}[t]
      \Vec{D}' & = \promote(i, x, \Vec{D})
    \end{aligned}
    \droptext{and}
    \begin{aligned}[t]
      \Vec{D}' & = \demote(i, x, \Vec{D})
    \end{aligned}
  \end{equation*}
\end{definition}

\begin{definition}[Collective Choice Function] Is a function that maps each
  situation to a subset of the feasible subsets for that situation. The
  collective choice function $F$ yields the \emph{choice set} of the
  \emph{proposal set}.
  \begin{equation*}
    \begin{aligned}[t]
      F \colon \mathcal{X} \times \powerSet{\Vec{D}} \mapsto \mathcal{X}
    \end{aligned}
    \quad
    \begin{aligned}[t]
      F(Y,\Vec{D}) \subseteq Y \\
    \end{aligned}
  \end{equation*}
  Typically the choice set will contain only one policy, the one that is
  preferred over all others. The chosen policy can then be removed from $Y$
  and the next policy found. In the event of a tie---if ties are
  tolerated---the choice set will contain the tied policies.
\end{definition}

\begin{remark}[Quorums] A quorum is usually taken to be the minimum number
  of voters that can demand that a policy be a legitimate choice. However,
  it may be that case that there is a quorum, but all votes, bar one, are
  abstentions and that the choice of that single individual becomes
  mandatory.

  For the time being, this anomaly should be noted, and there will be
  references in the text to the validity of a policy decision.
\end{remark}

\subsection{Conditions on a Collective Choice Function}
\label{sec:scf:conditions}

These conditions prescribe the behaviour of a \emph{collective choice
  function}\footnote{The more common term is social choice function, but
  this due to its origin in social welfare economics.}. These are derived
from Arrow's work\cite{inst:arrow} and have been the subject of
considerable debate. This rendering is original and, it is hoped, is more
explicit, self--contained and rigorous than that given in Arrow's work.
Each condition is expressed as a deduction rule in second--order logic: if
the premises are fulfilled then the conclusion is \emph{required} \ie it is
expected behaviour. The conditions therefore constitute tests for an
implementation of a collective choice function.

A more succinct rendering can be found in a paper by Batteau \etal
\cite{scf:batteau:theory}. It requires some familiarity with Arrow's
conditions and the theory of games\footnote{See for example,
  \cite{DN:games}.}, but has the advantage of relating Arrow's conditions
to work in games theory and also to requirements on the behaviour of
collective choice functions. This latter task has been carried out very
successfully by Fishburn\cite{scf:fishburn}, but only for two issue
collective choice functions. The paper by Batteau \etal only addresses
collective choice functions that use strong orderings.

There are five conditions in all. There is a brief description of the
meaning of each.

\begin{condition}[Admissible Orderings] This is a specification that the
  function need only operate on what are called \emph{admissible}
  orderings, an individual's ordering is admissible if it alone satisfies
  the collective choice function, \viz
  \begin{equation*}
    \frac{
      \begin{matrix}
        & \forall Y \in \mathcal{X} \\
        & \forall i \exists \Vec{D}
        [ \Vec{D} = ( \succ_i ), \varnothing \ne F(Y, \Vec{D}) \subset Y ]
      \end{matrix}
      } {
      \exists \Vec{D}' [
      \Vec{D}' = ( \succ_1, \dots, \succ_i, \dots, \succ_n ),
      \varnothing \ne F(Y, \Vec{D}) \subset Y
      ]
      } 
  \end{equation*}
  By specifying that individual voters must present orderings that are
  proper subsets of $Y$, this eliminates orderings that are completely
  indifferent or are cyclical, \eg $X = \{ x, y, z\}, \succ_i = (\ x > y,\ 
  y > z, \ z > x )$---so it is a condition on the vote set as well as on
  the collective choice function. In practice, it would be best to ensure
  that the orderings in $\Vec{D}$ are well--formed.
  
  The choice set $F()$ cannot be empty and it must be a proper subset of
  $Y$. If it is neither of these then there is either global indifference
  or a policy cycle.
\end{condition}

\begin{condition}[Monotonicity] or ``positive association of social and
  individual values''\cite[p. 25]{inst:arrow}: put simply if the
individuals want something and choose it for their society; if, in a later
vote, more individuals choose it, then, \textit{ceterus paribus}, it will
be chosen for society again, or, more formally:
  \begin{equation*}
    \frac{
      \begin{matrix}
        & \forall Y \in \mathcal{X} \\
        & \forall \Vec{D} \exists S [ F(Y, \Vec{D}) = S ] \\
        & \forall x \in S \exists \Vec{D}' [
        \Vec{D}' = \promote(x, \Vec{D})
        ]
      \end{matrix}
      } {
      \exists S' [ S' = F(Y, \Vec{D}'), x \in S' ]
      }
  \end{equation*}
\end{condition}

\begin{condition}[Independence] or ``independence of irrelevant
  alternatives'' \cite[p. 27]{inst:arrow} requires that the collective choice
  function return a choice set regardless of any individual's preferences
  for policies that are not explicitly part of the proposal set. This means
  that individuals may take on or discard values, or they may change their
  values regarding other matters, but these changes should not effect those
  values that have not changed. Formally, this can be expressed thus:
  \begin{equation*}
    \frac{
      \begin{matrix}
        & \forall Y \in \mathcal{X} \\
        & \exists X, X' 
        [ Y \subseteq X, \ Y \subseteq X', \ X \ne X']
      \end{matrix}
      }{ 
      \forall \Vec{D}, \Vec{D}'
      [\Vec{D} = \sincere(X),\  \Vec{D}' = \sincere(X'), 
      F(Y, \Vec{D}) = F(Y, \Vec{D}')]
      }
  \end{equation*}
  This condition on the implementation of the collective choice function is
  probably unimportant in practice; normally, the input to the collective
  choice function is $\Vec{D}$ which \emph{only} contains the preferences
  on the contents of $Y$, but as can be deduced from the discussion of the
  effect of an irrelevant alternative in the Borda preferendum, see table
  \ref{tab:choice:1}, these can affect a preference order.
\end{condition}

The following two conditions are more contentious. They are different from
the other conditions in that they need not be applicable to all issues and
there are two types of tests one can apply.

\begin{condition}[Non--imposition] There is no bias in the collective
  choice function that causes it, on some issues, to yield a choice set
  that is insensitive to voters' preferences.

  The first test is a unilateral test, \viz
  \begin{equation*}
    \frac{
      \begin{matrix}
        & \exists Y \in \mathcal{X} \\
        & \forall \Vec{D} \exists S [ S = F(Y, \Vec{D})] \\
        & \forall x \in S \exists \Vec{D}'
        [ \Vec{D}' = \demote(x, \Vec{D}) ]
      \end{matrix}
      }{
      \nexists S' [ S' = F(Y, \Vec{D}'), S' \cap S \ne \varnothing ]
      }
  \end{equation*}
  That is, it should be possible on a particular set of issues to construct
  a vote that does not return a particular policy for all votes that select
  that policy.
  
  The second test is used in the event of a tie between some policies to
  ensure that the collective choice function does not prefer one policy over
  the other.
  \begin{equation*}
    \frac{
      \begin{matrix}
        & \exists Y \in \mathcal{X} [ \#Y \ge 3 ] \\
        & \exists S [ \#S \ge 1, F(Y, \Vec{D}) = S ] \\
        & \forall y,z \in Y \setminus S [ y \ne z ]
      \end{matrix}
      } {
      \begin{matrix}
        & \forall w \exists Y' [ \neg ( w \in Y ), Y' = Y \cup \{ w \} ] \\
        & \exists \Vec{D}' [ \Vec{D}' = \sincere(Y') ] \\
        & \nexists S' [ S' = F( Y', \Vec{D}'),
        (y \in S' \wedge \neg (z \in S') ) \vee
        (z \in S' \wedge \neg (y \in S') ) ]
    \end{matrix}
    }
  \end{equation*}
  It might not be immediately clear from this formulation but this is an
  exact statement of the irrelevant alternative anomaly observed in the
  Borda preferendum, see table \ref{tab:choice:1}.
\end{condition}

\begin{condition}[Non--dictatorial] \label{def:scf:non-dictatorial} There
  is no one individual whose choice on some issues is always returned by
  the collective choice function, a dictator, nor is there any one
  individual who can reject some policies, a vetoer\footnote{Vetoer is a
    noun constructed solely for the purposes of this exposition.}.
  (Unfortunately, there is some contention about the use of the term ``one
  individual'', see the discussion following.)

  \begin{enumerate}
  \item Unilateral Tests
    
    These test whether it is possible to overcome a dictator's choice or a
    vetoer's rejection. The dictator or vetoer is placed in position $1$ for
    convenience.

    \begin{enumerate}
    \item Dictator
    \begin{equation*}
      \frac{
        \begin{matrix}
          & \exists Y \in \mathcal{X} \\
          & \exists S \exists \Vec{D} [
          \Vec{D} = ( \succ_1 ), \ S = F(Y, \Vec{D}) ] \\
          & \forall x \in S \exists \Vec{D}' [ \forall \succ_i [ i \ne 1,
          \succ_i = \demote(x, \succ_i)],\ \Vec{D}' = ( \succ_1, \dots,
          \succ_n ) ]
        \end{matrix}
        }{
        \nexists S' [ S' = F(Y, \Vec{D}'), S' \cap S \ne \varnothing
        }
    \end{equation*}
    
  \item Vetoer
    \begin{equation*}
      \frac{
        \begin{matrix}
          & \exists Y \in \mathcal{X} \\
          & \exists \Vec{D} \exists S^c [
          \Vec{D} = ( \succ_1 ), \ S^c = Y \setminus F(Y, \Vec{D}) ] \\
          & \forall x \in S^c \exists \Vec{D}' [ \forall \succ_i [ i \ne 1,
          \succ_i = \promote(x, \succ_i) ], \ \Vec{D}' = ( \succ_1, \dots,
          \succ_n ) ]
        \end{matrix}
        }{
        \nexists {S'}^c [ {S'}^c = Y \setminus F(Y, \Vec{D}'),
                        {S'}^c \cap S^c \ne \varnothing ] 
        }
    \end{equation*}
  \end{enumerate}
  
  \item Tie--Breaking Tests
    
    Subtler tests are those that are applied in the event of a tie.

    \begin{enumerate}
      
    \item Dictator
      
      The dictator is preferred in some way. In that, if the dictator
      changes allegiance, policy changes, but if anyone else it does not.

      \begin{equation*}
      \frac{
        \begin{matrix}
          & \exists Y \in \mathcal{X} [ \#Y \ge 3 ] \\
          & \exists S [ S \subset Y, \#S \ge 1 ] \\
          & \forall y [ y \in Y \setminus S ]
        \end{matrix}
        } {
        \begin{matrix}
          & \forall i \exists {\Vec{D}'}_i
          [ {\Vec{D}'}_i = \promote(i, y, \Vec{D}) ] \\
          & \forall i \exists S_i
          [ F(Y, {\Vec{D}'}_i) = S_i ] \\
          & \nexists j \forall i [ i \ne j, S_i \ne S_j ]
        \end{matrix}
        }
      \end{equation*}
    
    \item Vetoer
      
      \begin{equation*}
      \frac{
        \begin{matrix}
          & \exists Y \in \mathcal{X} [ \#Y \ge 3 ] \\
          & \exists S [ S \subset Y, \#S = 1 ] \\
          & \exists x [ x \in S ]
        \end{matrix}
        } {
        \begin{matrix}
          & \forall i \exists {\Vec{D}'}_i
          [ {\Vec{D}'}_i = \demote(i, x, \Vec{D}) ] \\
          & \forall i \exists S_i
          [ F(Y, {\Vec{D}'}_i) = S_i ] \\
          & \nexists j \forall i [ i \ne j, S_i \ne S_j ]
        \end{matrix}
        }
      \end{equation*}
    \end{enumerate}
  \end{enumerate}
  
  This definition states that ``no one individual'' can dictate a vote,
  which would seem to suggest that individuals can change their minds, but
  must do so \textit{en masse}. Arrow requires that a \emph{deciding
    set}\footnote{Also called a \emph{preventing set} in
    \cite{scf:batteau:theory} or a \emph{winning set} as defined in
    \cite{voting:isbell}. Some useful rules on winning sets are defined in
    the latter paper.} of voters must change their preferences. How many
  need to be in that set can only be determined by analyzing a vote.  It
  may appear that simple majority rule does not appear to meet this
  criteria, since, in a close result, any voter can invert the result,
  \emph{but} under simple majority rule on two issues, this can be
  dismissed, because every voter has exactly the same capability, therefore
  the simple majority becomes the decisive set.
  
  So, as it stands, this condition is still not accurately expressed, it
  should state that no subset of voters that is not a deciding set can
  change the outcome. The problem is that the deciding set cannot be known
  until the votes have been cast and counted.
\end{condition}


\section{Possibility and Impossibility Theorems} 
\label{sec:choice:arrow}

Arrow, who originally addressed the problem of the distribution of social
welfare, developed these theorems as general statements about a class of
functions which seek to combine hierachies of preference relations. Such
functions would be of great use in any field where a joint policy must be
formulated.  Clearly, that includes distributed computing and the
``globalization'' of local access rules to databases to form lattices of
information flow, \myRef{sec:gen-access}.

There are two results:

\begin{enumerate}
\item Possibility theorem for two--policy elections

  Such a collective choice function does exist for elections which have only a
  choice between two policies.

\item Impossibility theorem for elections having more than two policies
  
  There is, in general, no such collective choice function for elections having
  more than two policies.

\end{enumerate}

The idea behind the proof of the possibility theorem has already been given
in the discussion of deciding sets, condition
\ref{def:scf:non-dictatorial}, but the same idea is used in the proof of
the impossibility theorem.

\subsection{Conflict Resolution Mechanisms}
\label{sec:scf:decisions}

Arrow's proof for the impossibility theorem consists of analyzing how a
collective choice function can choose one preference over another. The notation
is as used in \myRef{sec:scf:notation}.

\begin{definition}[Unanimous Choice] If the voters unanimously agree that
  one policy is preferred over all others then that policy is chosen.
  \begin{equation*}
    \frac{
      \begin{matrix}
        & \forall Y \in \mathcal{X} \\
        & \forall i \exists x \in Y [
        \forall \Vec{D} [ \Vec{D} = ( \succ_i ) ], x \in F(Y, \Vec{D})
        ]
      \end{matrix}
      }{
      \nexists S \forall \Vec{D}' [
      \Vec{D}' = (\succ_1, \dots, \succ_i, \dots, \succ_n ),
      S = F(Y, \Vec{D}'), x \ni S
      ]
      }
  \end{equation*}
  Note that with this formulation it is possible to be indifferent to $x$,
  but one cannot oppose it\footnote{This is, in essence the \emph{Pareto
      principle} of social welfare. It can be stated as: ``social welfare
    is increased by a change that makes at least one individual better off,
    without making anybody else worse off\cite{econ:samuelson}.'' Clearly,
    if one abstains then one feels one is not going to be worse off if $x$
    is chosen. If one opposes $x$ then one would be worse off if $x$ were
    chosen.}. For example, if some voter has %
  $\succ = (x = y, y > z, x > z)$ then %
  $\{x, y\} = F(\{x, y, z\}, (\succ))$.
\end{definition}

\begin{definition}[Biased Choice] If pairs of voters contradict one another
  over a policy, one policy is chosen over the other.
  \begin{equation*}
    \frac{
      \begin{matrix}
        & \forall Y \in \mathcal{X} \\
        & \forall x',y' \in Y \exists i,j \in I 
        [ i \ne j, x' > y' \in \succ_i, y' > x' \in \succ_j ]
      \end{matrix}
      }{
      \forall \Vec{D} \nexists S [ \Vec{D} = ( \succ_i, \succ_j ),
      S = F(Y, \Vec{D}), y' \in S ]
      }
  \end{equation*}
  Note that this rule can only be applied pair--wise, it cannot be applied
  to the population as a whole.
\end{definition}

\begin{definition}[Unresolved Choice] If pairs of voters contradict one
  another over a policy, neither policy is chosen.
  \begin{equation*}
    \frac{
      \begin{matrix}
        & \forall Y \in \mathcal{X} \\
        & \forall x',y' \in Y \exists i,j \in I 
        [ i \ne j, x' > y' \in \succ_i, y' > x' \in \succ_j ]
      \end{matrix}
      }{
      \forall \Vec{D} \nexists S [ \Vec{D} = ( \succ_i, \succ_j ),
      S = F(Y, \Vec{D}), x', y' \in S ]
      }
  \end{equation*}
  This, too, may only be applied pair--wise.
\end{definition}

\subsection{Inadequacy of Conflict Resolution Mechanisms}

Unanimity, with abstentions, does not resolve any conflicts. If one
attempts to do so using one of the other two choice methods, one or more
of the conditions will be breached.

\begin{enumerate}
\item Biased Choice

  With two voters there is no majority decision, so the collective choice
  function must prefer:
  \begin{enumerate}
  \item Either a policy
  \item Or a particular voter's choice of policy
  \item Or randomly choose one policy
  \end{enumerate}
  
  The first two lead to an imposed policy or indicate a dictatorship,
  respectively; the latter has been suggested \cite{scf:zeckhauser}, but it
  is rather arbitrary.

\item Unresolved Choice

  This method allows anyone to act as a vetoer.

\end{enumerate}

One might think that one can improve the biased choice method so that it
decides in favour of whichever policy has a simple majority over the other.
Unfortunately, if one admits a third voter to make the biased choice
decisive, then one also allows that third voter to present a third policy
choice; in which case, one is attempting to choose between three policies,
which is the problem one is trying to solve.

If, in an attempt to overcome this, one requires that there always be more
voters than issues, then on some issues at least one voter will be a
dictator, or vetoer\footnote{This minority power is often said to be the
  cause of the instability of proportional representation parliaments.}.
The dictator, or vetoer, is acting as a ``Kingmaker''. It is possible to
stand this dilemma on its head (or feet) and use it as the basis for a
collective choice function as in \cite{scf:stcred}.

\subsection{Are the Conditions Reasonable?}
\label{sec:reasonable}

Hopefully, it should now be clear that it is not possible to construct a
collective choice function that satisfies all the conditions given above
simultaneously. That said, one can argue that the requirements on the
collective choice function's behaviour are too demanding.

\begin{enumerate}

\item Decisiveness
  
  Implicit in the definition of the collective choice function is that it is
  decisive.

\item The Non--imposition and Non--dictatorial Conditions
  
  These conditions come in two forms. The first form is unilateral and
  quite acceptable, although it may even be desirable that one particular
  voter has an absolute veto over a policy. The other form of the condition
  is only invoked in the event of a tie, where the requirement is that no
  policy or voter be preferred over another. This, it has been seen, is the
  pivotal difference between simple majority rule for a two policy vote and
  a three (or more) policy vote. In the two policy vote in the event of
  there only being one vote separating those for and those against, every
  voter is equally decisive and will be supported by the majority. In a
  close three (or more) policy vote, some voters can choose the minority
  position to force a tie.
  
  The issue is, again, the decisiveness of the collective choice function
  and would so close a result be acceptable.
  
\item Independence from Irrelevant Alternatives
  
  The preferendum, see table \ref{tab:choice:1}, and Condorcet pairings are
  both examples of collective choice functions that are non--dictatorial
  but are not free from the effects of irrelevant alternatives. This is
  very unfortunate, since an irrelevant alternative is any policy that is
  ranked lower than the collective winner, but is ranked higher than the
  collective winner by some voters. This would happen when the voters are
  assessing the policies with different underlying values which are more
  abstract.
  
  If one looks at the rankings made in table \ref{tab:choice:1}, it is
  clear that voter $k$ agrees with the others that $y>z$ but disagrees with
  them regarding the merits of both $y$ and $z$ over both of $w$ and
  $x$.

\item Monotonicity
  
  It can also be argued that monotonicity should be sacrificed to achieve a
  consensus. There is an attractive suite of collective decision functions
  called ``Kingmaker Trees''\cite{scf:stcred}. These do not demonstrate
  monotonic behaviour, but can be used to obtain a decision in the presence
  of sophisticated voting.

\end{enumerate}

\subsection{Are three (or more) policy collective choice functions usable?}

Whether the conditions Arrow imposed on collective choice functions are
reasonable only arises when the votes are close and that a sophisticated
voter would vote in such a way that a policy cycle arises. Three (or more)
policy votes \emph{can} give decisive results and it would be very useful
to have them resolve issues: can one therefore quantify how reliable a
collective choice is? How likely is it that an election has been subverted
by sophisticated voters, given the distribution of votes. Referring again
to table \ref{tab:choice:1}, if it were possible to compare the two sets of 
votes, impartially, an administrator would be able to make a better choice
of final policy. In which case, it would be better to introduce more
irrelevant alternatives to make a more certain choice.

Technologically, this is feasible. There are cryptographic algorithms which 
would allow vote sets to be cast secretly \cite{crypto:schneier} and these
could then be assessed by an impartial arbitrator to make the most
appropriate choice. The basis for that choice would be probabilistic and an 
example of a probabilistic criterion that could be employed is given next.

\section{Max--Min Probabilities in Condorcet Pairings}
\label{sec:max-min}

There is an interesting paper by Usiskin \cite{scf:usiskin}, which
quantifies the probabilities for Condorcet pairings. The paper addresses
the ``Voting Paradox'', but this is slightly misleading, it addresses the
organization of votes within committees. It covers the same ground as the
seminal works of Black and Farquharson \cite{scf:black,scf:farquharson},
which describe how committee procedures can be abused, if a policy cycle
exists.

In a committee procedure, if there is a policy cycle then for all those
voting: 
\begin{itemize}
\item $A>B>C$ has exactly $\frac{1}{3}$ of vote
\item as does $C>A>B$
\item and $B>C>A$
\end{itemize}

If the policies are voted on in pairs, then the order in which they are
introduced will determine which is chosen, \viz
\begin{itemize}
\item $( A \versusOp B ) \versusOp C = C \quad
  \because \ A \versusOp B = A$
\item $( C \versusOp A ) \versusOp B = C \quad
  \because \ C \versusOp A = C$
\end{itemize}

The question that Usiskin resolves is how much more popular than one
another can they be. In the example above, they all have probability of
beating, or of being beaten, of $\frac{2}{3}$.

Denote by $X_{i}$ a real--valued random variable that represents the
proportion of a simple majority vote received for policy $i$. The
probability of a simple majority vote having the outcome that $X_i > X_j$
will be: $P(X_i > X_j)$. A policy cycle will be revealed if the
probabilities for all pairs, $P(X_1 > X_2)$, $P(X_2 > X_3)$ and so on, for
$n$ policies is non--zero and, finally, $P(X_n > X_1)$ is also non--zero.
The maximum minimum value will represent how much more popular one policy
can be over another so that a policy cycle might still result.

If one then has at least one policy that beats another by an amount that is 
greater than this, then there can be no policy cycle.

\begin{theorem}[Arbitrary Random Variables] The maximum minimum value for 
  the joint probability distribution of a set of $n$ arbitrary random
  variables is given by:
  \begin{equation}
    \label{eq:scf:max-min}
    \max \left\{ \min \left[
        P(X_1 > X_2), \ldots, P(X_{n-1} > X_{n}), P(X_n > X_1)
      \right] \right\} =
    \frac{n-1}{n}
  \end{equation}
\end{theorem}

This is as one would expect, looking at the example given above, where each
of $A$, $B$ and $C$ had a probability of $\frac{2}{3}$ of beating the
other, if one of these had a greater probability than that, there could be
no policy cycle. In the example given above, were there were only three
voters, this would mean the voters unanimously agreed on one policy being
preferred over at least one other. This may not be the winning policy, it
would allow at least one policy to be eliminated, then a two--policy
vote can be taken.

This result is rather depressing but one can appreciate its intuitive
correctness, because it tells us that the more policies there are, the more
difficult it is to have one policy beating all others.

This case of arbitrary random variables does not help in understanding the
behaviour of sophisticated voters. (The probabilistic events would all be
conditioned by at least the previous result, \viz $P(X_i > X_{i+1} | X_i >
X_{i-1})$ and probably would need to be conditioned by all events.

However, Usiskin does present a result which could be used to interpret
sincere voting results. The election results $P(X_i > X_j)$ would be based
on $X_i, X_j$ being independent random variables
  
\begin{theorem}[Independent Random Variables] The maximum minimum value for 
  the joint probability distribution of a set of $n$ independent random
  variables is given by:
  \begin{align*}
    \max & \left\{ \min \left[ P(X_1 > X_2), \ldots, P(X_{n-1} > X_{n}),
        P(X_n > X_1)
      \right] \right\} = b(n) \\
    \droptext{where} & b(n+1) > b(n) \droptext{and} \lim_{n \rightarrow
      \infty} b(n) = \frac{3}{4}
  \end{align*}
\end{theorem}

This latter result is quite encouraging, because if there is an election
where at least one votes has a probability of greater than $\frac{3}{4}$
then no policy cycle can exist. 

Usiskin also demonstrates a method for formulating the function $b(n)$
and presents some upper and lower bounds.

\section{Some Collective Choice Functions}
\label{sec:scf:functions}

It was mentioned above, in the discussion of quorums, that a simple
majority rule collective choice would be valid even if only one voter
expressed a choice and all the others abstained. Simple majority rule is
just one of a number of collective choice functions that could be employed.
It is worthwhile just listing the collective choice functions.  These are
only for two policy votes and, because of Arrow's conditions, cannot be
extended to three (or more) policy votes, but they are insightful to the
acceptability of collective choices. This summary follows Fishburn,
\cite{scf:fishburn} and the details are contained in appendix
\ref{sec:scf}, but suffice to say that when only two policies, $x$ and $y$,
are under consideration ternary logic can be used with $x>y$ being $1$,
$y<x$ being $-1$ and $x=y$ $0$.  There are some diagrams that illustrate
the different types of voting rule and outcomes, figure
\ref{fig:voting-rules}

From the discussion above, \myRef{sec:scf:decisions}, a Pareto--optimal
collective choice function can also be specified, which is not in the
appendix, but is discussed in \cite{schobbens:preference} where it is
called ``unanimous with abstentions''.

\begin{table}[htbp]
  \begin{center}
    \begin{tabular}[left]{|r|p{2 in}|c|}
      \hline
      Rank & Rule & Paretian \\
      \hline
      1 & Specified Majority & Yes \\
      2 & Simple Majority & Yes \\
      3 & Specified Majority & No \\
      4 & Simple Majority & No \\
      \hline
    \end{tabular}
    \caption{Ranking of Binary Voting Rules}
    \label{tab:scf:rules}
  \end{center}
\end{table}

\begin{description}
\item[Simple Majority] if $\sign(\Vec{D}) > 1$ then $x>y$ is the collective 
  choice, see \eqref{eq:scf:smr}.
\item[Non--minority] if $\textbf{1}(\Vec{D}) > n/2$ then $x>y$ is the
  collective choice, see \eqref{eq:scf:nmr}. This is a special case of the
  next type of rule.
\item[Specified Majority] if $\textbf{1}(\Vec{D}) > \alpha n$ then $x>y$,
  where $\alpha$ is some pre--defined constant in range $(0,1)$, see
  \eqref{eq:scf:amr}.
\item[Absolute Majority] if $\textbf{1}(\Vec{D}) > \alpha n$ then
  $x>y$ and $y>x$ otherwise, where $\alpha$ is some pre--defined constant
  in range $(0,1)$, see \eqref{eq:scf:amr}.
\item[Absolute Special Majority] if $\textbf{1}(\Vec{D}) \leq
  \alpha n$ then $y>x$, see \eqref{eq:scf:asmr}.
\item[Pareto Majority: For] if $\textbf{-1}(\Vec{D}) = 0$ and
  $\textbf{1}(\Vec{D}) > 0$ then $x>y$.
\item[Pareto Majority: Indifference] if $\textbf{0}(\Vec{D}) > 0$ and
  $\textbf{-1}(\Vec{D}) = \textbf{1} = 0$ then $x=y$.
\end{description}

Absolute special majority is a variant of absolute majority (it is an
absolute majority of votes against) and absolute majority is just a variant
of specified majority where the complement of the policy is \emph{not}
installed if the required vote count is not reached and, as noted,
non--minority rule is a variant of \emph{Specified Majority Rule}. So all
of these can be replaced by that rule.

It is possible to define a \emph{Paretian quality} which can be added to
any voting rule and requires there is no dissenting vote. (Paretian
indifference can be thought of as a vote for $z \eqdef (x=y)$ as opposed to
$w = \eqdef (x>y) \vee (y>x)$, so it is actually a Paretian vote on a
different pair of issues: $z$ and $w$.)  The rules can be ranked in a
qualitative order of difficulty of attaining them, see table
\ref{tab:scf:rules}.

\section{Summary}

As information systems become more sophisticated they will be used to
support human decision--making. The prospect of constructing virtual
organizations based on how people interact is attractive: they could
potentially be more responsive---Miller describes an evolving information
processing organization, \cite{Miller:evolving}, which develops its
internal structure using a genetic algorithm. There are already some
prototypes, \cite{Hubermann:beehive,Hubermann:practice}, which share
information based on past usage.

This paper acts as a warning that collective decision--making is not
something to be taken lightly. Even the most sophisticated voting systems
can give rise to erroneous results, \myRef{sec:three-policy}. Sophisticated
voters making policy choices could give rise to systems falling into stasis
or being subverted to execute the wrong policies. If a discretionary access
control mechanism used to control the release of information from databases
were put under the control of collective choice functions and determined
access rights based on the criteria proposed by Bertino \etal,
\myRef{sec:gen-access}, it would almost certainly prove to be a vulnerable
system.

There is clearly a need for a more sophisticated architecture to deal with
access requests which can only be expressed in an enterprise modelling
language which would have components similar to that described by ISO in
their ODP, \cite{ODP:presc}. This type of information processing system
would need software agents acting on behalf of individuals to ensure that
their information is protected. This would necessarily be a probabilistic
analysis, based on how trustworthy potential information users are and it
may prove expedient to develop systems that are insured against loss or
provide degrees of surety, like those proposed by Neumann \etal
\cite{econ:inet:licensing}.

These information systems and their users would constitute an economy very
much like the everyday commercial world occupied by institutions,
corporations and people---only it would be faster, less resource wasteful
and, if information system designers integrate the safeguards before they
are used, safer.


\appendix

\section{ODP Enterprise Entities}
\label{sec:odp-enterprise}

These are Booch \cite{des:booch} class diagrams of the relationships that
exist between entities in a distributed processing system, or, indeed, any
organization as described in \cite[Enterprise Modelling
Language]{ODP:presc}.

\begin{enumerate}
\item Communities, see figure \ref{fig:ODP-enterprise-community}

  \begin{figure}[htbp]
    \begin{center}
      \includegraphics{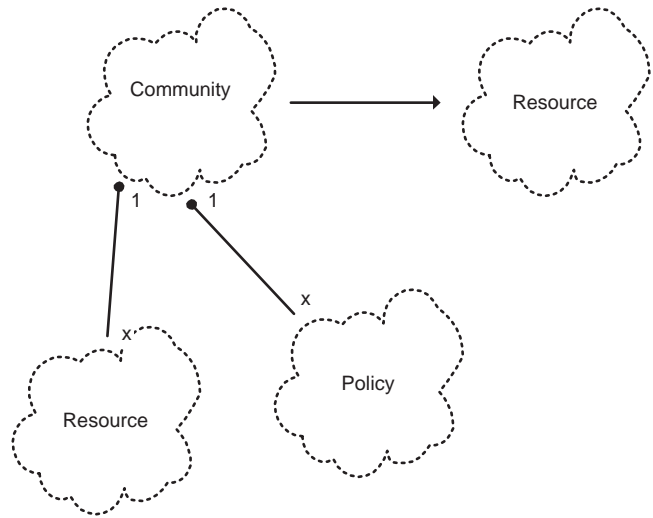}
      \caption{Communities, policies and resources}
      \label{fig:ODP-enterprise-community}
    \end{center}
  \end{figure}

  Communities comprise of collections of resources and policies. The
  community is itself a resource.

\item Enterprise Agents, see figure \ref{fig:ODP-enterprise-policies}

  \begin{figure}[htbp]
    \begin{center}
      \includegraphics{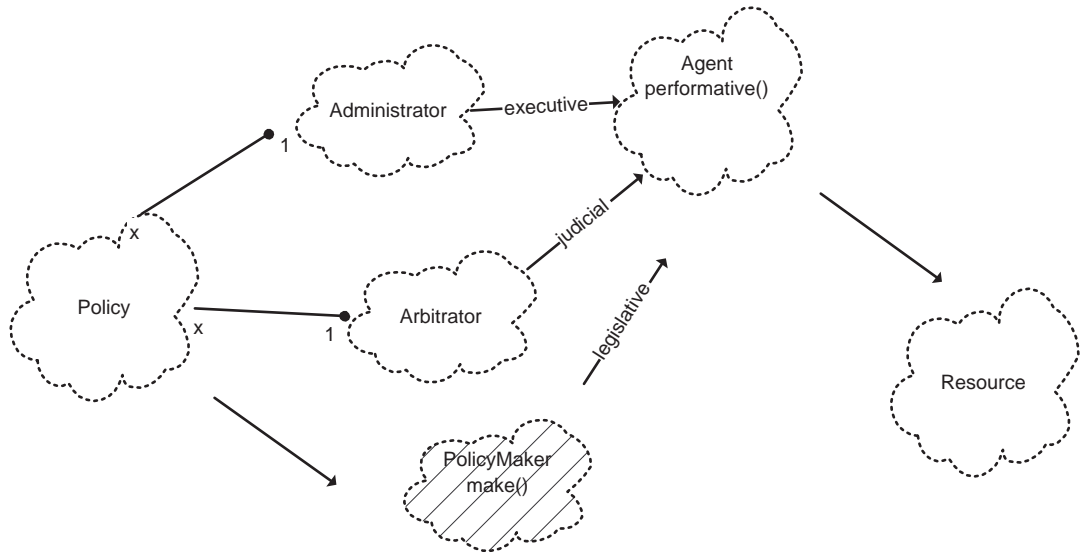}
      \caption{Agents (Policy--maker, Administrator, Arbitrator) and Policy}
      \label{fig:ODP-enterprise-policies}
    \end{center}
  \end{figure}
  
  Performative agents are also resources. There are three kinds of these:
  \begin{itemize}
  \item Administrators
  \item Arbitrators
  \item Policy--Makers
  \end{itemize}

  Administrators and arbitrators have policies they follow, but
  policy--makers create policy.

\item Resource Users
  
  These are also agents, but are not performative. They will have their own
  policies, but they are not explicitly open to arbitration. Resource users 
  may contact administrators prior to using a resource or they may not, it
  depends on the nature of the resource.
  
  Resource users usually control the policy--makers within communities and
  this is the case with most societies, since legislative assemblies are
  usually elected, but this need not be so. Companies are owned by its
  shareholders who appoint the board of directors, the administrators, but
  may not use the resources the company makes available. Suffice to say,
  that, in practice, in most business processes, the resource users have
  very little influence on the policy--makers.

\item Administrators, see figure \ref{fig:ODP-enterprise-resources}

  \begin{figure}[htbp]
    \begin{center}
      \includegraphics{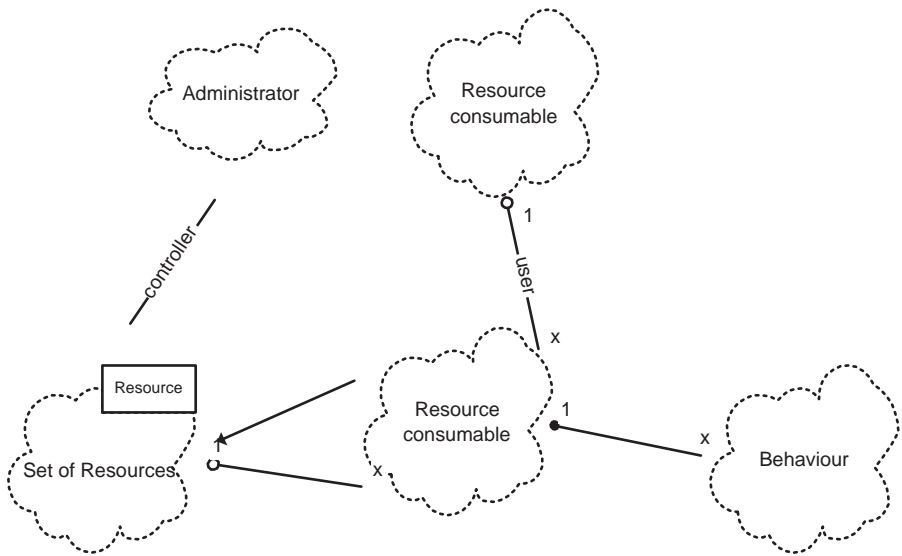}
    \caption{Administrators, Resources and Behaviour}
    \label{fig:ODP-enterprise-resources}
    \end{center}
  \end{figure}

  Administrators control sets of resources. Each consumable resource has a
  behaviour and may use other consumable resources. Since there may be
  different administrators vieing for the same consumable resources
  conflicts may arise.
  
  Resource users must obey the policy for using a resource. If there is no
  suitable policy, then the prospective user must have policy made. (This
  might seem different from what is observed in most organizations, where
  one can ask an administrator to apply policy differently in some way:
  usually by asking the administrator's superior to become involved. The
  administrator's superior is then acting as a policy--maker.)

\item Arbitrators, see figure \ref{fig:ODP-enterprise-control}

  \begin{figure}[htbp]
    \begin{center}
      \includegraphics{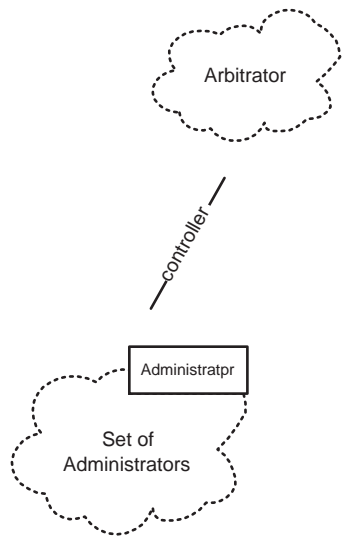}
    \caption{Arbitrators and Administrators}
    \label{fig:ODP-enterprise-control}
    \end{center}
  \end{figure}
  
  Arbitrators control adminstrators in that they resolve any conflicts that
  arise between them. They control neither administrators nor resources
  directly.

\item Policy, see figure \ref{fig:ODP-enterprise-statements}

  \begin{figure}[htbp]
    \begin{center}
      \includegraphics{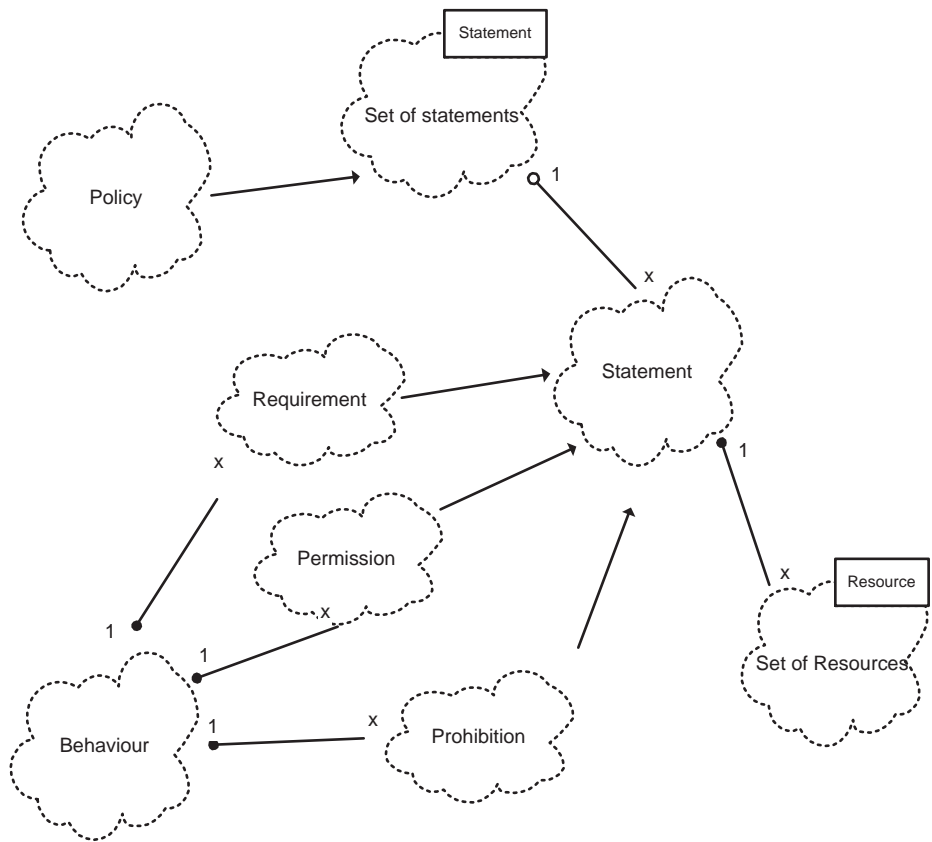}
    \caption{Policy, Resources and Behaviour}
    \label{fig:ODP-enterprise-statements}
    \end{center}
  \end{figure}
  
  Policy is a set of prescriptive statements about the behaviour of a set
  of resources: usually one sub--set of those resources \textit{vis \`a
    vis} another sub--set. There are three kinds of statement:
  \begin{itemize}
  \item Permission \- what one sub--set \emph{may} do with (or for) the
    other sub--set.
  \item Requirement (or Obligation) \- what one sub--set \emph{must} do
    with (or for) the other sub--set.
  \item Prohibition \- what one sub--set \emph{must not} do with (or for)
    the other sub--set.
  \end{itemize}

\end{enumerate}

\section{Collective Choice Functions}
\label{sec:scf}

What follows is drawn mostly from \cite{scf:fishburn} (who refers to proofs
from his own text\cite{scf:fishburn:1}: and gives a precise definition of
the conditions that a collective choice function must fulfill for a particular
voting procedure for two policy alternative systems.

\begin{definition}[Principle of Choice] The basic materials for collective
  choice functions are social alternatives (candidates, policies,
  \etc) and individuals (voters, members, \etc) who have preferences
  among the alternatives. The idea of a collective choice function is
  to map a non--empty subset of the potential feasible subset of
  alternatives to each ordered pair consisting of a potential feasible
  subset of alternatives and a schedule of the voters' preferences.
  The assigned set is often referred to as the \emph{choice set}.
\end{definition}

How that mapping is achieved is based on the properties of the collective
choice function, which decides whether the choice is:
\begin{itemize}
\item Egalitarian
\item Weighted 
\item Representative 
\item Unbiased (or neutral)
\item Decisive
\item Unanimous
\end{itemize}

\subsection{Two Policies}

If each individual has only two policies to choose from, then the policy
chosen by the population as a whole will always be one of them, so two
policy systems cannot select a set of policies that an individual
has not specified.
  
\begin{definition}[Sets for Two Policies] The sets can be enumerated
  quite easily for two policies $x, y$. If a voter prefers $x$ to $y$, then
  $1$ else if $y$ to $x$ then $-1$ else $0$ signifies
  indifference---ternary logic.
  \begin{displaymath}
    \begin{aligned}[t]
      \droptext{Policies} \\
      X = \{ x, y \} \\
      \mathcal{X} = \{ X \} \\
      \droptext{so write} \\
      F(\Vec{D}) \droptext{for} F(X, \Vec{D})
    \end{aligned}
    \quad
    \begin{aligned}[t]
      \droptext{$n$ voters} \\
      \Vec{D} = ( D_1, D_2, \dots, D_n ) \\
      \droptext{where} D_i \in \dset \\
      \droptext{so} \dset^n \eqdef \powerSet{\Vec{D}} 
      \droptext{and} \mathcal{\Vec{D}} \subseteq \dset^n
    \end{aligned}
  \end{displaymath}

  For the collective choice function over $n$ individuals:
  \begin{displaymath}
    F \colon \dset^n \mapsto \dset
  \end{displaymath}
\end{definition}

Note that the power set of the preferences is written $\dset^n$ as
shorthand and is the set of all permutations of vectors of length $n$ where
each component can take one of three values---$\#\dset^n = 3^n$. When a
condition is applied to a preference profile, it is either applied with
reference to the power set or the vote set: the power set, although large,
is denumerable \emph{a priori}, the vote set is not.

\subsection{Egalitarian}

Egalitarian\footnote{Fishburn in \cite{scf:fishburn} uses the term
  ``anonymous'' for egalitarian and ``dual'' for neutral.} collective choice
functions treat each voter's vote as identical in effect to every
other's.

\begin{condition}[\ref{eq:scf:neutral:0}] A collective choice function $F$ is
  \ref{eq:scf:neutral:0}, if, for all $\Vec{D} \in \dset^n$:
  \begin{equation}
    F(-\Vec{D}) = -F(\Vec{D})
    \tag{Strongly Neutral} \label{eq:scf:neutral:0}
  \end{equation}  
\end{condition}

\begin{condition}[\ref{eq:scf:smon}] A collective choice function $F$ is
  \ref{eq:scf:smon}, if, for any $\Vec{D}, \Vec{D}' \in
  \dset^n$:
  \begin{equation}
    \begin{split}
      \Vec{D} \geq \Vec{D}' & \Rightarrow F(\Vec{D}) \geq F(\Vec{D}') \\
      \Vec{D} > \Vec{D}', F(\Vec{D}') = 0 & \Rightarrow F(\Vec{D}) = 1
    \end{split} \tag{Strongly Monotonic} \label{eq:scf:smon}
  \end{equation}  
\end{condition}

\begin{condition}[\ref{eq:scf:egal}] A collective choice function $F$ is
  \ref{eq:scf:egal}, if for all $\Vec{D} \in \dset^n$:
  \begin{equation}
    \begin{split}
      F(D_1, \dots, D_n) & = F(D_{\sigma(1)}, \dots, D_{\sigma(n)}) \\
      \droptext{if} & \sigma \droptext{is a permutation on} \{1, \dots, n\}
    \end{split} \tag{Egalitarian} \label{eq:scf:egal}
  \end{equation}
\end{condition}

\begin{theorem}[Conditions for Simple Majority Rule\cite{scf:may:1}]
  \label{thm:scf:smf} A collective choice function $F$ implements simple majority
  rule over two policies and has the following qualities:
  \ref{eq:scf:neutral:0}, \ref{eq:scf:smon} and \ref{eq:scf:egal}.
\end{theorem}

\begin{definition}[Simple Majority Rule] If $F$ applies ternary logic, it
  can be implemented with:
  \begin{equation}
    \label{eq:scf:smr}
    F(D) \eqdef \sign(\Vec{D})
  \end{equation}
\end{definition}

\begin{condition}[\ref{eq:scf:mon}] A collective choice function is just
  \ref{eq:scf:mon}, rather than \ref{eq:scf:smon} if, for any $\Vec{D},
  \Vec{D}' \in \dset^n$:
  \begin{equation}
    \Vec{D} \geq \Vec{D}' \Rightarrow F(\Vec{D}) \geq F(\Vec{D}')
    \tag{Monotonic}
    \label{eq:scf:mon}
  \end{equation}
\end{condition}

\begin{definition}[Non--minority Rule] \label{eq:scf:nm} If a
  collective choice function is \ref{eq:scf:neutral:0},
  \ref{eq:scf:mon} and \ref{eq:scf:egal} and is implemented thus:
  \begin{equation}
    \label{eq:scf:nmr}
    \begin{split}
      F(\Vec{D}) = 1 & \Leftrightarrow \textbf{1}(\Vec{D}) > n/2 \\
      F(\Vec{D}) = -1 & \Leftrightarrow \textbf{-1}(\Vec{D}) > n/2
    \end{split}
  \end{equation}
  then the voting system is known as non--minority rule.
\end{definition}

\begin{figure}[htbp]
  \begin{center}
  \includegraphics{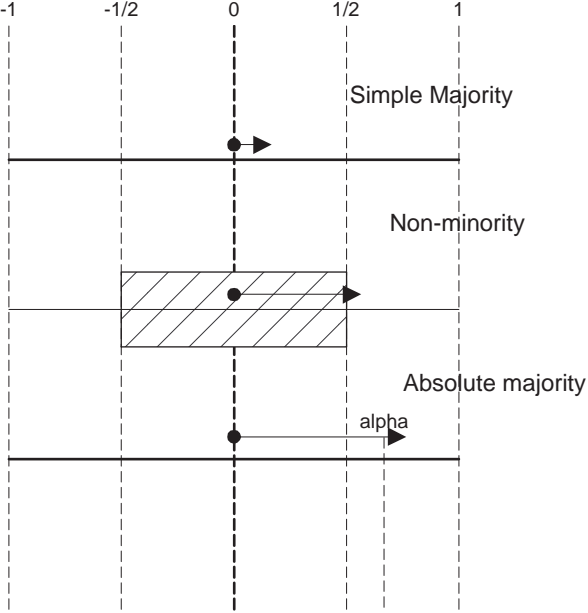}
  \caption{Two policy voting systems}
  \label{fig:voting-rules}
  \end{center}
\end{figure}

Non--minority rule is just one of a class of neutral, monotonic and
egalitarian collective choice functions; they differ in effect from
the strongly monotonic simple majority rule by having a
``dead--band''. A geometric insight into the specification of a
collective choice function can be given using a unit
simplex\cite{vote:saari}. There is only one dimension.
\begin{displaymath}
  \begin{split}
    \Vec{q} = (
    \frac{\textbf{1}(\Vec{D})}{n},
    \frac{\textbf{-1}(\Vec{D})}{n}
    )
  \end{split}
\end{displaymath}
The election vector $\Vec{q}$ emanates from the origin and will always
be within $[-1, 1]$. Under simple majority rule, whichever point, $-1$
or $1$, the vector is closest to wins. Under non--minority rule the
vector has to be over half $\frac{1}{2}$ way towards the point, see
figure \ref{fig:voting-rules}. The indecisive region in the centre is
symmetric. For non--minority rule, any boundary can be chosen, so long
as it is symmetric about the origin.

It should be clear from figure \ref{fig:voting-rules} that a simple
majority voting system can determine policy if only one voter is not
indifferent to either of the policies, the others abstaining. In this
respect, non--minority rule seems to impose a natural quorum, since it
requires that at least half of the voters have chosen one or the other
policy. In this respect, non--minority rule is less questionable as a
decision--making device than simple majority rule. Most electoral
systems do in fact operate a non--minority rule system.

\begin{condition}[\ref{eq:scf:neutral:1}] A collective choice function is
  \ref{eq:scf:neutral:1}, as opposed to \ref{eq:scf:neutral:0}, if, for all
  $\Vec{D} \in \dset^n$:
  \begin{equation}
    \textbf{1}(\Vec{D}) \ne \textbf{-1}(\Vec{D})
    \Rightarrow F(-\Vec{D}) = -F(\Vec{D})
    \tag{Neutral} \label{eq:scf:neutral:1}
  \end{equation}
\end{condition}

\begin{condition}[\ref{eq:scf:sdec}] A collective choice function is
  \ref{eq:scf:sdec} if, for all $\Vec{D} \in \dset^n$:
  \begin{equation}
    F(\Vec{D}) \ne 0
    \tag{Strongly Decisive} \label{eq:scf:sdec}
  \end{equation}
\end{condition}

\begin{condition}[\ref{eq:scf:unam}] A collective choice function is
  \ref{eq:scf:unam} if:
  \begin{equation}
    F(\Vec{1}) = 1 \droptext{and} F(\Vec{-1}) = -1
    \tag{Unanimity unambiguous} \label{eq:scf:unam}
  \end{equation}
\end{condition}

\begin{condition}[\ref{eq:scf:abst}] A collective choice function is
  \ref{eq:scf:abst} if, for all $\Vec{D} \in \dset^n$:
  \begin{equation}
    \begin{split}
      \droptext{if} \Vec{D} = \Vec{D}'
      \droptext{except that} & (D_i, D'_i) = (0,1) \\
      \droptext{for some $i$ then} & F(\Vec{D}) = F(\Vec{D}') \\
    \end{split}
    \tag{Pro--biased} \label{eq:scf:abst}
  \end{equation}  
\end{condition}

The number of electoral ties can be reduced by downgrading
\ref{eq:scf:neutral:0} to \ref{eq:scf:neutral:1} and adding
\ref{eq:scf:sdec}. If \ref{eq:scf:neutral:1} is dropped then an
\emph{electoral preference} is given to one policy over the other. The
policy that is preferred is usually already in force and is therefore
called the incumbent, the other policy is the challenging policy. See
figure \ref{fig:voting-rules} for the asymmetry of the indecisive region
under an absolute majority rule.

\begin{definition}[Absolute Majority Rule] If the collective choice
  function is no longer \ref{eq:scf:neutral:1} and is made
  \ref{eq:scf:sdec}, and the function is implemented thus:
  \begin{equation}
    \label{eq:scf:amr}
    \begin{aligned}
    F(\Vec{D}) = 1 & \Leftrightarrow \textbf{1}(\Vec{D}) > \alpha n \\
    F(\Vec{D}) = -1 & \Leftrightarrow \textbf{1}(\Vec{D}) \leq \alpha n \\
    \droptext{for} & \alpha \in (0,1)
    \end{aligned}
  \end{equation}
  then these are the absolute majority rule functions. A special case is
  \emph{unaminous rule} which requires either all votes are $-1$ or $1$.
\end{definition}

\begin{definition}[Absolute Special Majority Rule] As for absolute majority 
  rule, but the function is also \eqref{eq:scf:unam} and
  \eqref{eq:scf:abst}. The collective choice function can be
  implemented by:
  \begin{equation}
    \label{eq:scf:asmr}
    \begin{aligned}
    F(\Vec{D}) = -1 & \Leftrightarrow \textbf{1}(\Vec{D}) \leq \alpha n \\
    \droptext{for} & \alpha \in (0,1)
    \end{aligned}
  \end{equation}
\end{definition}

\subsection{Non--Egalitarian: Weighted}

The effect of the \ref{eq:scf:egal} property is that one voter's preference
can be exchanged for another within the decision profile and it will have
no effect on the evaluation of the collective choice function. An alternative
system is to use a weighted system, which is often used by some committees,
where the chairman is given both a deliberative and a casting
vote. Weighted systems have a number of attractions because they can be
designed so that:
\begin{itemize}
\item No one person can dictate policy to the group.
\item Ties can be readily resolved without a further vote.
\end{itemize}

The simplest way to define a weighted voting function is to use a weighting
vector.

\paragraph{Weighted Voting}

\begin{definition}[Weighting Vector and Vote] A weighting vector can be
  defined thus:
  \begin{equation*}
    \begin{split}
      \myVec{\rho}{n} > \myVec{0}{n},
      & \Vec{\rho} = ( \rho_1, \dots, \rho_n) \\
      \droptext{where} \rho_i \ge 0,\ \rho_i \in \field{Z}^{+}_0
      & \droptext{is the weight assigned to each voter}
    \end{split}
  \end{equation*}
  Then
  \begin{equation*}
    \begin{split}
      \droptext{Weighted Vote} & \Vec{\rho} \cdot \Vec{D} \\
      \droptext{Weight Function} & W(\Vec{\rho}) \eqdef \sum_{i=1}^{n} a_i \\
      \droptext{Weight of} c \in \dset &
      W_{c}(\Vec{D}) \eqdef
      W( (\Vec{\rho} \cdot \Vec{D}) : D_i = c) =
      \sum_{i=1}^{n} \rho_i D_i \droptext{for those} D_i = c
    \end{split}
  \end{equation*}
\end{definition}

It does \emph{not} follow that $n(\Vec{\rho}) \ge n$ because it is possible
to set any number of $\rho_i$ equal to zero, but at least one must be
non--zero. 

\begin{theorem}[Weighted Majority Function] A function $F$, $F \colon
  \dset^n \mapsto \dset$, is a weighted majority function if and only if it
  satisfies \ref{eq:scf:mon}, \ref{eq:scf:unam} and \ref{eq:scf:neutral:1},
  the weighted majority function can be defined as
  $F(\Vec{rho}\cdot\Vec{D}) = \sign(\Vec{\rho} \cdot \Vec{D})$.
\end{theorem}

The conditions can be readily tested by substituting $F(\Vec{\rho} \cdot
\Vec{D})$ for $F(\Vec{D})$.

The other systems can be added thus:
\begin{enumerate}
\item Non--minority rule

  \begin{displaymath}
    F(\Vec{\rho} \cdot \Vec{D}) = 
    \begin{cases}
      1 & W_{1}(\Vec{\rho} \cdot \Vec{D}) > \frac{W(\Vec{\rho})}{2} \\
      -1 & W_{-1}(\Vec{\rho} \cdot \Vec{D}) > \frac{W(\Vec{\rho})}{2} \\
      0 & \droptext{Otherwise}
    \end{cases}
  \end{displaymath}

\item Absolute majority

  \begin{displaymath}
    F(\Vec{\rho} \cdot \Vec{D}) = 
    \begin{cases}
      1 & W_{1}(\Vec{\rho} \cdot \Vec{D}) > \alpha W(\Vec{\rho}) \\
      -1 & W_{-1}(\Vec{\rho} \cdot \Vec{D}) \leq \alpha W(\Vec{\rho}) \\
      0 & \droptext{Otherwise}
    \end{cases}
  \end{displaymath}

\end{enumerate}

\paragraph{Dictators and Vetoers}

The final choice may be wholly determined by only one of the voters, in
which case that voter is either a \emph{dictator} or a \emph{vetoer}.

\begin{enumerate}

\item Dictator

\begin{definition}[\ref{eq:scf:dict}] A voter, $j$, is a \ref{eq:scf:dict}
  with regard to a collective choice function $F$ and a weighting vector
  $\Vec{\rho}$ if:
  \begin{equation}
    \begin{split}
      \droptext{For all} & \Vec{D} \in \dset^n \droptext{such that} \\
      \Vec{D} & = ( D_1, \dots, D_j, \dots, D_n) \\
      \droptext{when} &
      D_j \ne 0, \ F(\Vec{\rho} \cdot \Vec{D}) = D_j
    \end{split}
    \tag{Dictator} \label{eq:scf:dict}
  \end{equation}
\end{definition}

Whether a dictator can effect \emph{all} decision profiles is a
condition on the behaviour of the collective choice function and the weighting
vector, not on the voters, which is as follows.

\begin{definition}[\ref{eq:scf:ndom}] A collective choice function is
  \ref{eq:scf:ndom} if there is \emph{no} \ref{eq:scf:dict}. This can be
  stated thus:
  \begin{equation}
    \begin{split}
      \droptext{For all} & \Vec{D} \in \mathcal{\Vec{D}} \\
      \droptext{There is no $i$ such that} & \\
      \Vec{D} & = ( D_1, \dots, D_i, \dots, D_n), \\
      D_i \ne 0 \droptext{and} F(\Vec{D}) = D_i
    \end{split}
    \tag{Undominated by dictator} \label{eq:scf:ndom}
  \end{equation}
\end{definition}

This is not particularly useful, since there may legitimately be a voter
whose vote is always in line with the choice of the group as a whole.

\item Vetoer

\begin{definition}[\ref{eq:scf:vetr}] The first voter, $1$, is a
  \ref{eq:scf:vetr} with regard to a collective choice function $F$ and a
  weighting vector $\Vec{\rho}$ if:
  \begin{equation}
    \begin{split}
      \droptext{For} & \\
      \Vec{D}_1 & = ( 0, 1, \dots, 1) \\
      \Vec{D}_{-1} & = ( 0, -1, \dots, -1) \\
      F(\Vec{\rho} \cdot \Vec{D}_1) = 0 \\
      F(\Vec{\rho} \cdot \Vec{D}_{-1}) = 0
    \end{split}
    \tag{Vetoer} \label{eq:scf:vetr}
  \end{equation}
  (The vetoer is in first position for convenience).
\end{definition}

Whether a vetoer can effect those two very specific decision profiles is a
condition on the behaviour of the collective choice function and the weighting
vector, not on the voters, which is as follows.

\begin{definition}[\ref{eq:scf:ndom:1}] A group of voters are
  \ref{eq:scf:ndom:1} if there is no voter $j$, such that:
  \begin{equation}
    \begin{split}
    \droptext{If} & \\
    D_j = 1,& \ F(\Vec{\rho} \cdot \Vec{D}_{1}) = 1, \droptext{and} \\
    D_j = -1,& \ F(\Vec{\rho} \cdot \Vec{D}_{-1}) = -1 \\
    \droptext{but} & \\
    D_j \ne 1,& \ F(\Vec{\rho} \cdot \Vec{D}_{1}) = 0 \\
    D_j \ne -1,& \ F(\Vec{\rho} \cdot \Vec{D}_{-1}) = 0
    \end{split}
    \tag{Undominated by vetoer} \label{eq:scf:ndom:1}
  \end{equation}
  For all $\Vec{D} \in \mathcal{\Vec{D}}$.
\end{definition}

This, again, is not particularly useful, since there may legitimately be a
voter whose always votes against the group.

\end{enumerate}

\paragraph{Sensitivity}

\begin{definition}[\ref{eq:scf:esse}] A voter $i$ is said to be 
  \ref{eq:scf:esse} with regard to a collective choice function, and weighting
  vector, if at least one of the following conditions holds:
  \begin{equation}
    \begin{split}
      \droptext{Either} F(\Vec{\rho} \cdot \Vec{D}_{1}) & \ne F(\Vec{\rho}
      \cdot \Vec{D}_0) \\ 
      \droptext{Or} F(\Vec{\rho} \cdot \Vec{D}_{-1}) & \ne F(\Vec{\rho}
      \cdot \Vec{D}_0) \\ 
      \droptext{Or} F(\Vec{\rho} \cdot \Vec{D}_1) & \ne F(\Vec{\rho} \cdot
    \Vec{D}_{-1})
    \end{split} \tag{Essential} \label{eq:scf:esse}
  \end{equation}
  For at least one of the vectors $\Vec{D}_1, \Vec{D}_0, \Vec{D}_{-1}$
  constructed as follows:
  \begin{align*}
    \Vec{D}_1 & = ( D_1, \dots, D_{i-1},1,D_{i+1}, \dots, D_n) \\
    \Vec{D}_0 & = ( D_1, \dots, D_{i-1},0,D_{i+1}, \dots, D_n) \\
    \Vec{D}_{-1} & = ( D_1, \dots, D_{i-1},-1,D_{i+1}, \dots, D_n)
  \end{align*}
  where the contents of those vectors can be taken from any of the vectors
  constructed thus:
  \begin{equation*}
    \begin{split}
      \myVec{D}{n-1} & = ( D_1, \dots, D_{i-1}, D_{i+1}, \dots, D_n ) \in
      \dset^{n-1}
    \end{split}    
  \end{equation*}
\end{definition}

This condition requires that a voter \emph{can} be decisive in at least
one decision profile. It prevents a voter from being given so ineffectual a
vote that it is never decisive in any election.

\subsubsection{\ref{eq:scf:vetr}, \ref{eq:scf:dict} and \ref{eq:scf:esse}}

By this it is meant safe from dictators and vetoers and sensitive to
voters; it is desirable if a collective choice function and weighting vector
could be chosen so that for all decision profiles in $\dset^n$ there is no
voter who is either a \ref{eq:scf:dict} or a \ref{eq:scf:vetr}. It would
also be desirable there is at least one voter who is \ref{eq:scf:esse}.

The collective choice functions simple majority, non--minority and absolute
majority are demonstrably safe from dictators and vetoers when used under
an egalitarian regime so only the weighting vector needs to be checked.

\paragraph{Dictators, Vetoers and Weighting Vectors}

\begin{enumerate}

\item Weighted majority and weighted non--minority rule
  
  If the collective choice function is either of the above, for
  $\Vec{\rho} = ( \rho_1, \dots, \rho_n )$ and the weighting vector
  has been scaled so that $\rho_1$ has weight $1$. In which case, the
  worst case for the dictator is that all vote against, so for the
  dictator to succeed $\rho_{dictator} > \frac{W(\Vec{\rho})}/2$.

  Because both of these collective choice functions are dual; it should be
  clear $\rho_{dictator} = \rho_{vetoer}$.

  Consequently, $\rho_{max} < \frac{W(\Vec{\rho})}{2}$, is sufficient for
  both of these.

\item Absolute majority

  To succeed, $\rho_{dictator} > \alpha \cdot W(\Vec{\rho})$, so the
  converse is required.

  A vetoer has it easier $\rho_{vetoer} > (1 - \alpha) \cdot
  W(\Vec{\rho})$, so the converse.

\end{enumerate}

\paragraph{Weighting Vectors that are \ref{eq:scf:esse} to Voters}

Under egalitarian rule, all voting functions are sensitive to all voters,
because if any voter is sensitive, a permutation can put another voter in
his place, so again the collective choice function will not be at fault should
a system of rule prove insensitive, it will be the weighting vector.

There are two possibilities:
\begin{itemize}
\item The voter has a weighting of zero
\item The voter has a weighting which can never be decisive
\end{itemize}

Whether a voter has a non--zero vote can only be tested for; probably by
using the weighted majority rule function with all other voters not voting.

Having a vote that is never decisive is more subtle. The weighting vector
is as usual, with the lowest rated voter in first position having value
1. Construct a dictator to each voter in the following manner.

\begin{align*}
  \rho_1 = 1 \\
  \rho_2 = \rho_1 + 1 \\
  \rho_3 = \rho_2 + \rho_1 + 1 \\
  \dots \\
  \rho_{n-1} = \rho_{n-2} + \dots + \rho_2 + \rho_1 + 1 \\
  \rho_n = \rho_{n-1} + \rho_{n-2} + \dots + \rho_2 + \rho_1 + 1 \\
  \rho_n = \rho_{n-1} + \rho_{n-1} \\
  \rho_n = 2 \rho_{n-1} \\
  \rho_n = 2 \cdot 2 \cdot \cdot \droptext{$n - 1$ times} \cdot 1 \\
  \rho_n = 2^{n - 1}
\end{align*}

A necessary conditions on weighting vectors can be set, each $\rho_i <
2^{n-1}$ if $W(\Vec{\rho}) \ge 2^n - 1$---geometric progression. There are
effectively two choices:
\begin{itemize}
\item Set $n \ge 3, \ \max{\rho_i} = 2^{n - 1} - 1$, in which case
  $W(\Vec{\rho}) \ge 2^n - 2$ and all other voters will tie with the
  largest voter.
\item Set $n \ge 4, \max{\rho_i} = 2^{n -1} - 2$, in which case
  $W(\Vec{\rho}) \ge 2^n - 3$ and all other voters will defeat the
  largest voter.
\end{itemize}

\begin{align*}
  \droptext{For} \Vec{\rho} = ( \rho_1, \dots, \rho_i, \dots, \rho_n ) \\
  \nexists \rho_i > \frac{1}{2} \sum_{j=1}^{n} \rho_j 
\end{align*}

\subsection{Representative Systems}

A representative system can be thought of as a heirachy of voting councils
in which the outcomes of votes in lower councils become votes in higher
councils. A voter in one of the higher councils may be a voter or a voting
council. Each voting council can use weighted majority rule between its
members. Voters (or councils) can vote more than once in different
councils.

\begin{definition}[Representation] Let a heirachy of voting councils be
  defined as a \emph{representation} $\mathcal{R}$. Let each level of
  representation be denoted by a suffix, the lowest level being
  $\mathcal{R}_0$.
  
  To make the lowest level similar in mathematical structure to higher
  levels, we shall introduce a selection function $S_i$ which selects from
  a preference profile, $D$, the preference of voter $i$.
  \begin{align*}
    S_i \colon \dset^n \mapsto \dset \\
    S_i(D) = D_i    
  \end{align*}
  
  $\mathcal{R}_0$ can then be written as:
  \begin{displaymath}
    \mathcal{R}_0 = \{ S_1(D), \dots, S_n(D) \}
  \end{displaymath}
  So $\mathcal{R}_0$ is simply the decision profile as a set.

  Thereafter, there is a level $m \in \field{N}$ which is such that:
  \begin{displaymath}
    \mathcal{R}_m = \{ \sign(F_1, \dots, F_K)(\mathcal{R}_{m-1}) \}
  \end{displaymath}
  
\end{definition}

This is effectively voting using a tree structure and is probably the
most used organizational control system. Unfortunately, it is proving
to be very difficult to analyze. Hopefully, more results will arise.


\section{Funding and Author Details}

Research was funded by the Engineering and Physical Sciences Research
Council of the United Kingdom. Thanks to Malcolm Clarke, Russell--Wynn
Jones and Robert Thurlby.

\begin{verse}
  Walter Eaves \\
  Department of Electrical Engineering, \\
  Brunel University \\
  Uxbridge, \\
  Middlesex UB8 3PH, \\
  United Kingdom
\end{verse}

\begin{verse}
  \url{Walter.Eaves@bigfoot.com} \\
  \url{Walter.Eaves@brunel.ac.uk}
\end{verse}

\begin{verse}
  \url{http://www.bigfoot.com/~Walter.Eaves} \\
  \url{http://www.brunel.ac.uk/~eepgwde}
\end{verse}

\sloppy
\newcommand{\etalchar}[1]{$^{#1}$}
\hyphenation{ Kath-ryn Ker-n-i-ghan Krom-mes Lar-ra-bee Pat-rick Port-able
  Post-Script Pren-tice Rich-ard Richt-er Ro-bert Sha-mos Spring-er The-o-dore
  Uz-ga-lis }

\end{document}